\documentclass[aps,prb,superscriptaddress,twocolumn,floatfix,showpacs,a4paper]{revtex4-1}
 \usepackage{amsfonts,amsmath,amssymb}
\usepackage{textcomp}
\usepackage{bm}
\usepackage{graphicx}
\usepackage[latin1]{}
\usepackage{color}

\newcommand{\notop}{{{}_{}}}

\newcommand{\ie}{\textit{i.e.}}

\newcommand{\etal}{\textit{et~al.}}

\renewcommand{\vec}[1]{\bm{#1}}
\newcommand{\ee}{\mathrm{e}}

\newcommand{\pp}{\partial^{{}}}

\newcommand{\nablabf}{\boldsymbol{\nabla}}

\newcommand{\rot}{\nablabf\times}

\newcommand{\calH}{\mathcal{H}}

\newcommand{\AAA}{\vec{A}}

\newcommand{\GGG}{\vec{G}^\notop}

\newcommand{\HHH}{\vec{H}}

\newcommand{\KKK}{\vec{K}}

\newcommand{\ppp}{\vec{p}}

\newcommand{\rrr}{\vec{r}}

\newcommand{\uuu}{\vec{u}}

\newcommand{\eps}{\epsilon}

\newcommand{\ket}[1]{|#1\rangle}
\newcommand{\braket}[2]{\langle #1 | #2 \rangle}

\newcommand{\beq}[1]{\begin{equation} \eqlab{#1}}
\newcommand{\eeq}{\end{equation}}
\newcommand{\bsub}{\begin{subequations}}
\newcommand{\esub}{\end{subequations}}
\def\bal#1\eal{\begin{align}#1\end{align}}
\def\bsubal#1\esubal{\bsub \begin{align}#1\end{align} \esub}

\newcommand{\eqlab}[1]{\label{eq:#1}}
\renewcommand{\eqref}[1]{Eq.~(\ref{eq:#1})}
\newcommand{\eqsref}[2]{Eqs.~(\ref{eq:#1}) and~(\ref{eq:#2})}

\newcommand{\figref}[1]{Fig.~\ref{fig:#1}}

\newcommand{\secref}[1]{Section~\ref{sec:#1}}

\newcommand{\seclab}[1]{\label{sec:#1}}

\begin{document}

\title{Pseudomagnetic fields and triaxial strain in graphene }

\author{Mikkel Settnes} \email[]{mikse@nanotech.dtu.dk}
\author{Stephen R. Power}
\author{Antti-Pekka Jauho}
 \affiliation{Center for Nanostructured Graphene (CNG), DTU Nanotech, Technical University of Denmark, DK-2800 Kongens Lyngby, Denmark}

\date{\today}

\begin{abstract}
	Strain fields in graphene giving rise to pseudomagnetic fields have received much attention due to the possibility of mimicking real magnetic fields with magnitudes of greater than 100 Tesla.
	We examine systems with such strains confined to finite regions  (``pseudomagnetic dots") and provide a transparent explanation for the characteristic sublattice polarization occurring in the presence of pseudomagnetic field. 
	In particular, we focus on a triaxial strain leading to a constant field in the central region of the dot. 
	This field causes the formation of pseudo Landau levels, where the zeroth order level shows significant differences compared to the corresponding level in a real magnetic field. 
	Analytic arguments based on the Dirac model are employed to predict the sublattice and valley dependencies of the density of states in these systems.
	Numerical tight binding calculations of single pseudomagnetic dots in extended graphene sheets confirm these predictions, and are also used to study the effect of the rotating the strain direction and varying the size of the pseudomagnetic dot. 
	\end{abstract}

\maketitle
 
\section{Introduction}
One of the many remarkable properties of graphene is the close connection between electronic structure and mechanical deformation. Non-uniform deformations introduce massive amplitude pseudomagnetic fields (PMFs) \cite{CastroNeto2009,Vozmediano2010} within the effective Dirac approximation.
An intriguing consequence of a uniform PMF is the development of Landau quantization in the absence of magnetic fields \cite{Guinea2009,Neek-Amal2013,Guinea2010}. 
The dramatic impact of moderate deformations has lead to the concept of strain engineering \cite{Pereira2009,Guinea2012,Jones2014} which suggests using the PMF to manipulate the valley degree of freedom in graphene \cite{Chaves2010,Low2010,Zhenhua2011,Kim2011,Qi2013} or to introduce electronic band gaps. \cite{Ni2009,Low2011} 

Experiments by Levy \etal \; \cite{Levy2010} and Lu \etal \; \cite{Lu2012} on nanometer sized graphene bubbles have demonstrated pseudo Landau levels (pLLs) corresponding to a PMF with magnitude exceeding several hundreds of Tesla. 
Different approaches have since been suggested to control the applied local strain via direct pressure from STM tips \cite{Klimov2012}, gas-inflation \cite{Bunch2008,Georgiou2011,Zenan2014,Settnes2015,Bahamon2015} or substrate structuring/interaction \cite{Tomori2011,Reserbat-Plantey2014,Gill2015,Scharfenberg2011,Zang2013,Bao2009,Meng2013,Neek-Amal2012c,Sloan2013,Shioya2014}. 
In this way, strain engineering holds the promise of very localized and strong PMFs together with the possibility of continuous tunability. 
In this work, we focus on the effect of the PMF on the local density of states and employ an analytical form of the strain field from which an analytical PMF can be easily derived.
The formation and magnitude of strain-induced PMFs can be studied in more detail using molecular dynamic approaches \cite{Zenan2014,Neek-Amal2012a,Neek-Amal2012b,Jones2014}. 

Motivated by the experiments discussed above showing pLLs arising in nano-sized bubbles, we focus on such local strain fields (``pseudomagnetic dots") embedded within infinite graphene sheets.
We consider PMFs which give rise to pLLs in pseudomagnetic dots and provide a simple explanation within the Dirac approximation for the special sublattice polarization observed in such PMF systems \cite{Schneider2015,Neek-Amal2013,Carrillo-Bastos2014,Moldovan2013,Poli2014,Settnes2015}. This analysis is furthermore confirmed by numerical tight binding calculations using the recently developed patched Green's function approach \cite{Settnes2015}, which allows us to calculate the local electronic structure without the necessity to assume a finite sample, or to introduce periodicity.
This method is used to further investigate the interplay between strain direction and sublattice polarization, and finite-size effects which occur in small pseudomagnetic dots.


This paper is organized as follows. \secref{TB_strain} briefly reviews how lattice deformations are included within the standard tight binding model and also the low energy Dirac Hamiltonian, leading to the introduction of PMFs. 
\secref{analyticA} analyses general sublattice polarization in a PMF and \secref{psDot} introduces the applied model for a finite pseudomagnetic dot supporting pLLs. 
\secref{results} presents numerical results confirming our analysis and discusses further the effect of the strain direction and the size of the pseudomagnetic dot with respect to the resulting pLL structure.

\section{Strain within tight-binding models} \seclab{TB_strain}
We treat graphene through a nearest neighbor tight binding Hamiltonian 
\bal 
\calH = \sum_{<i,j>} t_{ij} c_i^\dagger c_j, \eqlab{TB_H}
\eal
where the sum $<i,j>$ runs over nearest neighbor sites.
For a pristine graphene sheet, this model is characterized by a constant carbon-carbon hopping matrix element $t_0=-2.7\mathrm{eV}$. 
When the atoms are displaced relative to each other the bond lengths vary leading to a spatially dependent hopping integral, $t_{ij} = t(\rrr_i,\rrr_j)$.
The position of an atom is given by $\rrr_i = \rrr_i^0 +\uuu$ where $\rrr_i^0$ is the equilibrium position and $\uuu(x,y)=(u_x(x,y),u_y(x,y),z(x,y))$ is the displacement field. 
In equilibrium the bond length is $a_0=0.142$ nm but after displacement it changes to $d_{ij}=|\rrr_i-\rrr_j|$ and the hoppings $t_{ij}$ are correspondingly modified according to the commonly used model\cite{Pereira2009,Moldovan2013,Carrillo-Bastos2014}
\bal 
t_{ij}= t_0 \ee^{-\beta\big(d_{ij} /a_0 -1\big)}, \eqlab{t_strain}
\eal
where $\beta = \pp \log(t) / \pp \log(a)|_{a=a_0}\approx 3.37$. \cite{Pereira2009}

The new bond length can also be obtained from the strain tensor \cite{LandauBook}
\bal \eqlab{dij_strain}
d_{ij} = \frac{1}{a_0} \big(a_0^2 + \eps_{xx}x_{ij}^2 + \eps_{yy}y_{ij}^2 + 2\eps_{xy}x_{ij}y_{ij}\big), 
\eal	
where $\rrr_{ij} = \rrr_i-\rrr_j$ and the strain tensor is given by classical continuum mechanics as \cite{Vozmediano2010} 
\bal \eqlab{strain_tensor}
\eps_{ij} = \frac{1}{2} \bigg(\partial_j u_i +\partial_i u_j + (\partial_i z)(\partial_j z)\bigg), \quad i,j = x,y.
\eal
The low energy effective Dirac Hamiltonian for deformed graphene then takes the form \cite{CastroNeto2009,Goerbig2011} 
\bal
H_{\KKK}(\ppp) =   v_F \bm{\sigma} \cdot \bigg(\ppp +  e\AAA \bigg), \eqlab{Hdirac_strain}
\eal
where $\ppp$ is the momentum measured from $\KKK$, ${\bm \sigma} = [\sigma_x, \sigma_y]$ with $\sigma_{x/y}$ being the usual Pauli matrices, $v_F$ is the pristine Fermi velocity and for simplicity we have omitted explicit reference to the band index. 
The Hamiltonian for the $\KKK'$ valley is obtained using the transformations ${\bm \sigma} \rightarrow -{\bm \sigma}$ and $\AAA \rightarrow -\AAA$  such that time reversal symmetry is conserved. We note that intrinsically there is no coupling between the valleys within this low-energy model.

The effective gauge field $\AAA$ in \eqref{Hdirac_strain} is given by the two dimensional strain tensor $\eps_{ij}(x,y)$ \cite{Suzuura2002,Jones2014,Faria2013}
\bal \eqlab{Afield}
\AAA =-\frac{\hbar\beta}{2e  a_0} \left( \begin{matrix}
	\eps_{xx}-\eps_{yy} \\
	-2\eps_{xy}
\end{matrix} \right). 
\eal
We note that \eqref{Afield} only takes into account the first order corrections in the hopping parameter. Expanding to higher orders in the deformation leads to Fermi surface anisotropy \cite{Pereira2010,Pellegrino2010} and spatially dependent Fermi velocity \cite{Manes2013,Ramezani2013,deJuan2013,Jang2014}. 
Analogously to a real vector potential, the strain-induced vector potential generates a so-called pseudo magnetic field (PMF), $B_s$, perpendicular to the graphene sheet \cite{Guinea2009,Vozmediano2010}. 
The sign of the PMF depends on the valley such that, in the $\KKK$ valley, the field is given by 
\bal \eqlab{Bs}
B_s &= \rot \AAA = \partial_x A_y - \partial_y A_x, 
\eal
whereas the opposite sign is taken in the $\KKK'$ valley because $\AAA\rightarrow -\AAA$.
Importantly, the definition of the pseudomagnetic field is inherently connected to a first order expansion of \eqsref{TB_H}{t_strain} in the low energy Dirac model of graphene. We stress, however, that the numerical calculations presented below for electronic properties are based on a full tight binding model with hopping parameters given by \eqref{t_strain}.

\section{Pseudomagnetic dots with Landau levels} \seclab{analytic}
If we restrict ourselves to a single valley, the gauge field in \eqref{Afield} enters the Dirac Hamiltonian in the same way as a real magnetic field does  (see \eqref{Hdirac_strain}). 
We can therefore compare a uniform PMF to a real magnetic field.
In the presence of a real magnetic field the electronic spectrum is modified giving rise to Landau quantization \cite{Goerbig2011}. However, as opposed to conventional (non-relativistic) Landau levels which have a spectrum linear in the $B$-field, the Landau levels in graphene follows a characteristics $\sqrt{Bn}$-behavior including a zero energy Landau level ($n=0$). 
The analogy between real and pseudomagnetic fields therefore suggests the existence of pseudo Landau levels in the presence of a constant pseudomagnetic field \cite{Guinea2009,Levy2010},
\bal
E_n = \mathrm{sign}(n) \sqrt{2e \hbar v_F^2 B_s |n|}, \eqlab{En_LL}
\eal
where $E_n$ is the energy for the Landau level $n$. 
The corresponding magnetic length $l_B$ is given by the usual expression
\bal 
l_B = \sqrt{\frac{\hbar}{e  B_s}}.\eqlab{lB} 
\eal
 Indeed, signatures of such levels have been detected using scanning tunneling microscopy to explore the spatial distribution of states in strained bubbles of nanometer size formed on graphene \cite{Levy2010,Lu2012,Li2015}. 

\subsection{Sublattice polarization } \seclab{analyticA}
The conservation of time reversal symmetry, leading to an opposite sign of the PMF in opposite valleys, has interesting consequences for the sublattice occupation in the presence of a PMF.
The solution to the two dimensional Dirac Hamiltonian around $\KKK$ is a two dimensional spinor $\ket{\Psi_{\KKK}} = \big(\psi_{\KKK}^\bullet,\psi^\circ_{\KKK}\big)^T$, where $\bullet$ denotes the A sublattice and $\circ$ denotes the B sublattice. The spinor components of valley $\KKK'$ satisfy the same type of Dirac equation as $\KKK$ so that both valleys can be conveniently collected in a four component spinor $\ket{\Psi}$ containing both valleys, \cite{Beenakker2008,Goerbig2011}
\bal
\ket{\Psi}  =
 \begin{pmatrix}
	\psi_{\KKK}^\bullet \\\psi_{\KKK}^\circ \\   \psi_{\KKK'}^\circ \\ \psi_{\KKK'}^\bullet
\end{pmatrix}, \eqlab{Psi4dim}
\eal
following the notation of Refs. \onlinecite{Goerbig2011,CastroNeto2009} and noting that the sublattice sequencing is reversed between the valleys. 
This definition gives rise to a four-component Dirac Hamiltonian with two unequal subblocks
\bal
\begin{pmatrix}
	v_F {\bm \sigma} \cdot (\ppp + e\AAA) &	\mathbf{0} \\
	\mathbf{0} & -v_F {\bm \sigma} \cdot (\ppp - e\AAA)\\
\end{pmatrix} \ket{\Psi} = E \ket{\Psi}. \eqlab{H4dim}
\eal
However, we note that other representations using a valley isotropic formulation with two equal blocks are also common in the literature \cite{Beenakker2008}. 

We first conclude from \eqsref{Psi4dim}{H4dim} that interchanging the valley indices inverts the role of the two sublattices, \ie \; $\big(\psi_{\KKK}^\bullet ,\psi_{\KKK}^\circ \big) \rightarrow \big( \psi_{\KKK'}^\circ ,\psi_{\KKK'}^\bullet \big)$. 
Second, we note that the transformations $\AAA \rightarrow -\AAA$ and $\ppp \rightarrow -\ppp$ interchanges the role of the subblocks in \eqref{H4dim}.
A solution of the two-dimensional Dirac equation in \eqref{Hdirac_strain}, for the $\KKK$ valley and positive magnetic field, can be written in the form $\ket{\Psi_{\KKK}} = c_1 \ket{\KKK, \bullet} + c_2 \ket{\KKK, \circ}$, where $c_1$ and $c_2$ are coefficients determining the wavevector component on the A ($\bullet$) and B ($\circ$) sublattices, respectively, for the solution in a real magnetic field. \footnote{We note that the $c_1$ and $c_2$ coefficients contain sign factors depending on both valley and band index \cite{Goerbig2011}. In the discussions of \secref{analyticA}, however, we omit these factors for simplicity since we focus on features like the LDOS which only depend on the magnitude of the wavefunction.}
Considering this form of the solution, we use the symmetries discussed above to determine the form of the wavefunction for different valleys and signs of the $B$-field: 
\bal 
\begin{array}{c |c| c}
	&B_s  & -B_{s}\\ [0.15cm] \hline  
	\ket{\Psi_{\KKK}}   &  c_1 \ket{\KKK, \bullet} + c_2 \ket{\KKK, \circ} &  c_2 \ket{\KKK, \bullet} + c_1 \ket{\KKK, \circ}\\ [0.15cm]   \hline
	\ket{\Psi_{\KKK'}} &   c_2 \ket{\KKK', \bullet} + c_1 \ket{\KKK', \circ} & c_1 \ket{\KKK', \bullet} + c_2 \ket{\KKK', \circ} \\ [0.15cm] \hline	
\end{array} \eqlab{BK_matrix}
\eal
First, we consider the total wavefunction in the presence of a {\it real} magnetic field, $\ket{\Psi^{re}}$, which has the same sign of the $B$-field in both valleys,
\bal 
\ket{\Psi^{re}} =\begin{pmatrix}
	\ket{\Psi_{\KKK}}_B \\  \ket{\Psi_{\KKK'}}_B
\end{pmatrix} 
\eal
From the first column of \eqref{BK_matrix} we can determine the sublattice dependency of the wavefunction in the presence of a real magnetic field. The contribution to the density of state on the $\bullet$ sublattice from the $\KKK$ valley is expressed as $\rho^\bullet_{\KKK} \cong \braket{\psi^{\bullet}_{\KKK}}{\psi^{\bullet}_{\KKK}}$.
The total density of states on the $\bullet$ sublattice is proportional to a sum of contributions from each valley, $\rho^\bullet \cong \rho^\bullet_{\KKK} + \rho^\bullet_{\KKK'} = |c_1|^2 + |c_2|^2$. 
A similar argument holds for the $\circ$ sublattice using
\bsub
\bal
\braket{\psi^{\bullet}_{\KKK}}{\psi^{\bullet}_{\KKK}}_B &= \braket{\psi^{\circ}_{\KKK'}}{\psi^{\circ}_{\KKK'}}_{B} =|c_1|^2,  \\
\braket{\psi^{\circ}_{\KKK}}{\psi^{\circ}_{\KKK}}_B &= \braket{\psi^{\bullet}_{\KKK'}}{\psi^{\bullet}_{\KKK'}}_{B} =|c_2|^2 . 
\eqlab{rho_polarization_expl_real}
\eal
\esub
We note that although the sublattices have identical LDOS distributions, the contributions to these from the individual valleys are swapped. A consequence of this is that the individual sublattices are valley polarized for the zeroth Landau level in a real magnetic field. \cite{Goerbig2011}
This is equivalent to $c_1=0$ in the above notation, which results in sublattice $\bullet$ being occupied only by $\KKK'$-valley fermions and similarly the $\circ$ sublattice occupied entirely by the $\KKK$-valley.

Returning to the case of pseudomagnetic fields, we can express the total wavefunction, $\ket{\Psi^{ps}}$, by components from $\KKK$ and $\KKK'$ with opposite sign of the $B$-field, corresponding to the diagonal elements of \eqref{BK_matrix}, 
\bal 
\ket{\Psi^{ps}} = \begin{pmatrix}
	\ket{\Psi_{\KKK}}_B \\  \ket{\Psi_{\KKK'}}_{-B}
\end{pmatrix} 
\eal
Using \eqref{BK_matrix}, the density of states for each sublattice is proportional to
\bsub \eqlab{rho_polarization_expl}
\bal
\braket{\psi^{\bullet}_{\KKK}}{\psi^{\bullet}_{\KKK}}_B &= \braket{\psi^{\bullet}_{\KKK'}}{\psi^{\bullet}_{\KKK'}}_{-B} =|c_1|^2,  \\
\braket{\psi^{\circ}_{\KKK}}{\psi^{\circ}_{\KKK}}_B &= \braket{\psi^{\circ}_{\KKK'}}{\psi^{\circ}_{\KKK'}}_{-B} =|c_2|^2 . 
\eal
\esub
Here we note that we get the same contribution to the LDOS in a given sublattice from both valleys. 
This is in contrast to the situation in a real magnetic field, where each valley gives a different contribution to the LDOS within a single sublattice.
In a PMF, the overall sublattice equivalence is thus broken when we consider the contributions from both valleys combined -- the LDOS in sublattice A is given by $\rho^\bullet \sim 2 |c_1|^2$ and that in sublattice B by $\rho^\circ \sim 2 |c_2|^2$.
This gives rise to the characteristic sublattice polarization in PMF's noted in Refs. \onlinecite{Settnes2015,Neek-Amal2013,Moldovan2013,Schneider2015}.
The above analysis reveals that this is a general feature arising due to pseudomagnetic fields causing the two valleys to experience equal but opposite $B$-fields.

\subsubsection*{Zeroth Landau level in a pseudomagnetic field}
The sublattice polarization in the presence of a PMF has a special consequence for the DOS distribution of the zeroth order pseudo Landau level. 
As the two-dimensional form of the Dirac equation for the $\KKK$ valley in a PMF is identical to that for a real field, we have $c_1=0$.
From \eqref{rho_polarization_expl} we then conclude that the $n=0$ pLL is entirely confined to the B ($\circ$) sublattice, with a density of states proportional to $|c_2|^2$ and an equal contribution from each valley.
This sublattice polarization has been noted in theoretical works \cite{Neek-Amal2013} examining PMFs in hexagonal graphene flakes but here it arises as an entirely general feature disconnected from the presence of edge features or particular geometries.
The sublattice polarization is in contrast to the $n=0$ Landau level in a real magnetic field, where both sublattices contribute to the density but are completely valley polarized. \cite{Goerbig2011}

\begin{figure}[htb]
	\begin{center}
		\includegraphics[width= 0.40\columnwidth]{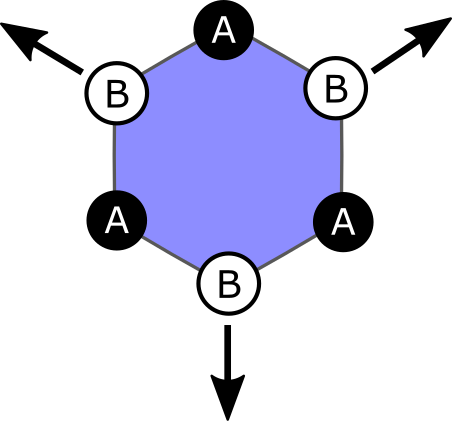}
		\caption[]{ \small Schematic representation of the triaxial strain in \eqref{triaxial_u_ac} relative to the crystal orientation. 
		} \label{fig:strain_sketch}
	\end{center}
\end{figure}

\subsection{Finite size pseudomagnetic dots} \seclab{psDot}
In the following, we focus on finite strained regions (``dots") within an infinite sheet of graphene motivated by the experimental realizations of graphene bubbles. \cite{Levy2010,Lu2012} 
We apply the triaxial strain field suggested by Guinea \etal \; \cite{Guinea2009} as this gives rise to a constant PMF qualitatively similar to the actual experimental strain field exhibiting pLL features. The displacement field is given by
\bal \eqlab{triaxial_u_ac}
\uuu(r,\theta) & = \begin{pmatrix}
	u_x \\u_y
\end{pmatrix} = \begin{pmatrix}
u_0 r^2 \sin(3\theta) \\ u_0 r^2 \cos(3\theta)
\end{pmatrix},
\eal
where $(r,\theta)$ give the positions of the atoms in polar coordinates. In this way, \eqref{triaxial_u_ac} gives rise to a strain along the armchair direction as sketched in \figref{strain_sketch}. 
From \eqref{Afield} we conclude that \eqref{triaxial_u_ac} leads to a constant PMF given by $B_s =  8 u_0\frac{\hbar\beta }{2e a_0}$, where $u_0$ determines the strength of the strain field \cite{Guinea2009,Juan2011}. 

We note that Ref. \onlinecite{Neek-Amal2013} concluded that the effect of the PMF on the LDOS is qualitatively unchanged after relaxation by molecular dynamics. 
We therefore use \eqref{triaxial_u_ac} to directly relate an analytical, continuous form of the PMF to a discrete strain field on the graphene lattice. This can be used to parameterize a tight binding Hamiltonian for numerical calculations using \eqref{t_strain}. 

\begin{figure}
	\begin{center}
		\includegraphics[width= 0.95\columnwidth]{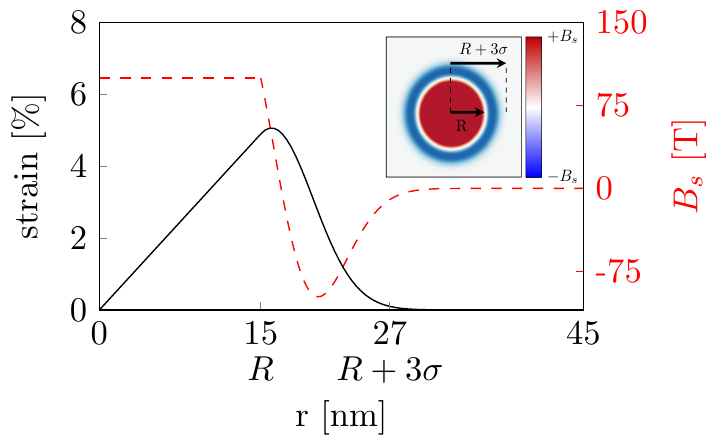}
		\caption[]{ The strain (black curve) and magnitude of the pseudomagnetic field (dashed curve) as a function of distance from the center of the pseudomagnetic dot with $R=15$ nm. The strain is taken along the armchair direction using the definition $\frac{d_{ij}-a_0}{a_0}$ where $d_{ij}$ is the new bond length and $a_0$ is the pristine bond length. The inset shows a map of the full PMF in the dot. This PMF distribution is generic for all different sizes of the pseudomagnetic dots, only the maximum strain will change for different dot sizes at same magnitude of the PMF.  
		} \label{fig:PMFdot}
	\end{center}
\end{figure}

To ensure a finite strained region, we apply a smoothing to the strain field in \eqref{triaxial_u_ac} using the transformation ${\bm \eps} \rightarrow {\bm \eps}\;\ee^{{-(r-R)^2/2\sigma^2}}$ for $r>R$, where $R$ is the radius of the constant PMF region. 
The resulting strain field is shown in \figref{PMFdot} for a 15 nm sized dot together with the resulting PMF. 
The inset in \figref{PMFdot} shows the full PMF distribution calculated using \eqsref{Afield}{Bs}. 
We clearly see a constant PMF for $r<R$, with a varying PMF of opposite sign for $r>R$ within the smoothing region. 
Note that this opposite sign field within the smoothing region arises because we apply the smoothing to the strain tensor and {\it not} to the PMF itself. 
The constant PMF, $+B_s$, for $r<R$ is caused by the {\it positive} change in the strain. 
However, the smoothing gives rise to a negative change in the strain which in turn leads to a PMF of opposite sign in the region $r>R$.
This sign change is a general feature and is also seen in studies using molecular dynamics to examine finite strained regions \cite{Zabet2014,Zenan2014}.
 

\section{Numerical calculations} \seclab{results}
We calculate the LDOS at every site within the strained region based on the full tight binding Hamiltonian using the real-space patched Green's function approach \cite{Settnes2015}. 
The Hamiltonian of the entire system is replaced by an effective Hamiltonian, $\HHH_{\mathrm{eff}} = {\bf \mathcal{H}} + {\bf \Sigma}_B$, whose dimension is the number of sites within the finite `patch' surrounding the pseudomagnetic dot.
In this form, $\calH$ describes the effect of the finite strain field and ${\bf \Sigma}_B$ contains the influence of the surrounding pristine graphene sheet on the strained region. 
This allows us to treat the area containing pseudomagnetic dot explicitly and thereby express the Green's function of the strained region as
\bal 
\GGG(E) = \big(E - \calH - {\bf \Sigma}_B \big)^{-1}.
\eqlab{GF_patch}
\eal
The boundary self-energy ${\bf \Sigma}_B$ ensures that we treat an infinite graphene sheet without any edges, allowing us to separate effects arising to due to the PMF with those that arise due to edges of the sample. 
We can express ${\bf \Sigma}_B$ conveniently using the pristine Green's functions along the boundary of the calculation area \cite{Settnes2015}
and these can be determined efficiently by exploiting complex contour techniques \cite{Power2011}. 
The boundary self-energy enters only along the edge of the calculation area and we can treat the retarded Green's function in \eqref{GF_patch} through a recursive method adapted to account for this extra self-energy as detailed in Ref. \onlinecite{Settnes2015}.
The local density of states is finally obtained from the imaginary part of the Green's function as $\rho_i(E) = -\mathrm{Im} \big[G_{ii}(E)\big]/\pi$.
In the results presented below, the density of states will be shown averaged over all the sites on a given sublattice in the central region of a pseudomagnetic dot, e.g. for $r<R/2$.

\subsection{Pseudo Landau levels in density of states}
\begin{figure}
	\begin{center}
		\includegraphics[width= 0.90\columnwidth]{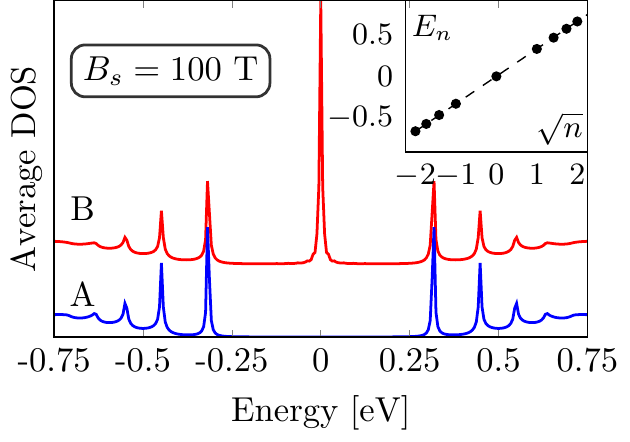}
		\caption[]{ \small Average density of states for both the A and B sublattice at the center of a region with radius of $R = 10$ nm subjected to a triaxial strain corresponding to $B_s = 100$ T. The smoothing region has a width $\sigma = 4$ nm and the curves are translated vertically with respect to each other.   
		The inset shows the positions of the pLLs plotted as a function of $\textrm{sign(n)}\sqrt{|n|}$ where the dashed curve is based on \eqref{En_LL} using $B_s=100$ T.
		} \label{fig:ldos_inplane_AB}
	\end{center}
\end{figure}

\begin{figure}
	\begin{center}
		\includegraphics[width= 1\columnwidth]{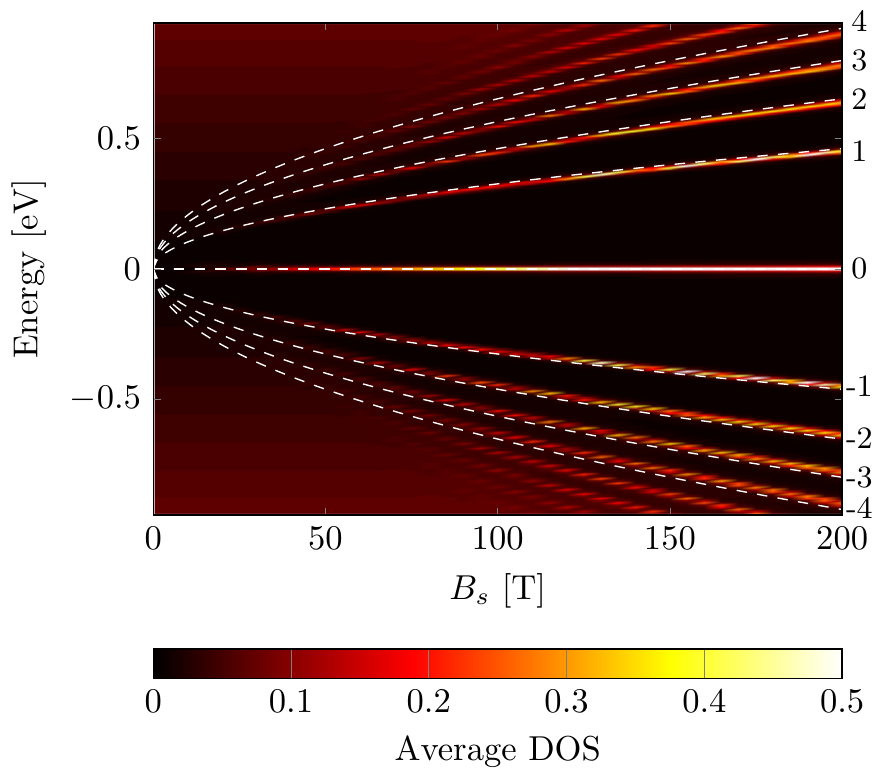}
		\caption[]{ \small A contour plot of the average DOS for $r<R/2$ ($R=15$ nm) as a function of the energy and the magnitude of the PMF. The white dashed curves are based on \eqref{En_LL} with the peak number indicated.
		} \label{fig:scan_map}
	\end{center}
\end{figure}

We consider a pseudomagnetic dot whose strain field, calculated from \eqref{triaxial_u_ac} corresponds to a PMF of $B_s=100$ T at the dot center. 
The average DOS within the central region of the dot ($r<R/2$) is shown in \figref{ldos_inplane_AB}, and the direction of the applied strain is shown in the inset. 
We notice from the inset of \figref{ldos_inplane_AB} that the appearance of peaks in the DOS is following the $\sqrt{B|n|}$ behavior predicted by \eqref{En_LL}, which is shown by the dashed line. 
These pLLs are well formed for lower energies, but tend to wash out and broaden at higher energies. This is clearly seen when considering the contour plot in \figref{scan_map} where the average DOS is indicated by the colormap as a function of energy and magnitude of the PMF. For a given magnitude of the PMF we see how only the first couple of pLLs are visible. In addition, the levels at higher fields tend to deviate more from the analytical predictions.
Furthermore, from \figref{ldos_inplane_AB} we note an important difference between these pLL's compared to regular Landau levels: the zeroth pLL only has a finite contribution to the LDOS on one sublattice, as also noticed in Refs. \onlinecite{Neek-Amal2013,Qi2013}, but not investigated in direct relation to the strain field.

This sublattice polarization of the zeroth pLL numerically confirms the analytical considerations discussed in \secref{analyticA}.
The sublattice with zero contribution is determined by the vanishing coefficient (either $c_1$ or $c_2$), which in turn is determined by the sign of the PMF. 
The solution yielding $c_1=0$ assumes a positive $B$-field in the $\KKK$. 
If this valley experienced a negative $B$-field, we would get a vanishing contribution for the opposite sublattice. 
The strain direction shown in the inset of \figref{ldos_inplane_AB} gives rise to a PMF with positive sign in the $\KKK$ valley, and thus a zeroth order pLL with finite contribution on the B sublattice.
This special connection between the sublattices, the zero order pLL and the direction of the triaxial strain is investigated in more detail below. 

\subsection{Rotation of triaxial strain}
\begin{figure}
 	\begin{center}  
 		\includegraphics[width= 0.35\columnwidth]{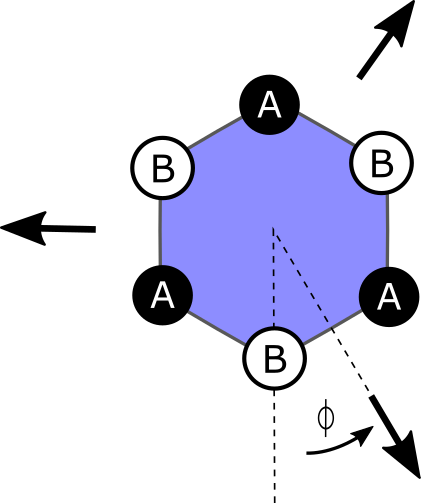}
 		\begin{minipage}{1\columnwidth}
 			\begin{center}\includegraphics[width=0.85\columnwidth]{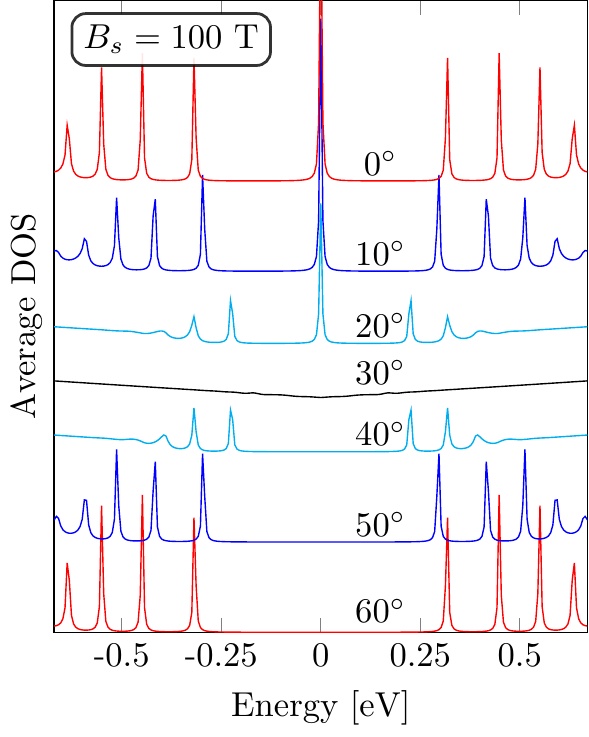}
 				\end{center}
 		\end{minipage} 
 		\caption[]{The average density of states for sublattice B at the center of a triaxial strain corresponding to $B_s=100$ T and a radius $R=15$ nm. The different curves correspond to different rotational angles $\phi$, and are translated with respect to each other for clarity.  
 		} \label{fig:ldos_inplane_varAngle}
 	\end{center}
 \end{figure}

The triaxial strain in \eqref{triaxial_u_ac}, giving rise to a constant pseudomagnetic field\cite{Guinea2009}, is along the high symmetry armchair directions of the graphene lattice.
We now consider a rotation of the triaxial strain by an angle $\phi$ from the armchair direction as illustrated by the top schematic in \figref{ldos_inplane_varAngle}. 
This leads to a generalised PMF of the form $B_s = B_0 \cos(3\phi)$ with a maximum amplitude when the strain is aligned with the armchair direction \cite{Verbiest2015}. 
On the other hand, a triaxial strain along the zigzag direction ($\phi = 30^\circ$) does not give rise to a PMF at all. 
We therefore do not observe any peak features in the low energy spectrum, even if the magnitude of the displacement field is the same but simply rotated to the zigzag direction. 

In \figref{ldos_inplane_varAngle} we show the average DOS on sublattice B as the rotation angle of the triaxial strain field is increased away from the armchair direction. 
First, we observe a decrease in the strength of the PMF as we rotate the strain field away from $\phi=0^\circ$ to $\phi=30^\circ$ . 
The decreasing PMF is evident from the lowering of the $n=1$ pLL as the angle is increased. Furthermore, we notice the very distinct Landau peaks for strain along the armchair directions ($\phi=0^\circ$ and $\phi=60^\circ$). Finally, the linear DOS observed at $\phi=30^\circ$ (zigzag direction) confirms the prediction of zero PMF for triaxial strain along the zigzag direction.
Clearly, the formation of a PMF is highly dependent on the direction of strain. This means that strain along multiple directions (e.g. rotationally symmetric strain) will lead to an inhomogeneous PMF where the pLLs will be less pronounced. 
For inhomogeneous PMFs the DOS may contain a combination of peaks making the identification of pLLs difficult in these systems, even before additional factors such as electron scattering near aperture edges for gas-inflated bubbles are taken into account \cite{Settnes2015,Settnes2015b,Zenan2014}. 

The results in \figref{ldos_inplane_varAngle} show the presence of a zeroth pLL peak for $\phi=0^\circ,10^\circ,20^\circ$ but not for $\phi=40^\circ,50^\circ,60^\circ$. 
The presented calculations do not contain any sample edges which can interact with the PMF. 
We therefore conclude that the sublattice polarization, and in particular which sublattice is occupied, is determined solely by the direction of the triaxial strain compared to the crystalline directions. 
A triaxial strain along the armchair direction can be applied in two different ways to the graphene lattice ($\phi=0^\circ$ or $\phi=60^\circ$ in \figref{ldos_inplane_varAngle}).
Using the analysis from \secref{analytic}, we note that the directions $\phi=0^\circ-30^\circ$ corresponds to $+B_s$ while the strain at $\phi=30^\circ-60^\circ$ corresponds to $-B_s$. Consequently, we have $c_1=0$ in \eqref{rho_polarization_expl} for $\phi=0^\circ-30^\circ$ and the zeroth order pLL therefore resides on the B ($\circ$) sublattice. On the other hand, for the directions $\phi=30^\circ-60^\circ$ the zeroth order pLL resides on the opposite sublattice.
It is also clear from the schematic in \figref{ldos_inplane_varAngle} that triaxial strain along the armchair direction, $\phi=0^\circ$, breaks sublattice symmetry by displacing atoms on the $A$ and $B$ sublattices differently. 
The other armchair possibility, $\phi=60^\circ$, simply inverts the roles of the sublattices.
Meanwhile, all triaxial strains along zigzag directions are sublattice symmetric as they are perpendicular to bonds connecting sites on opposite sublattices and thus affect these sites equally.
The intrinsic connection between sublattice asymmetry and pseudomagnetic fields is emphasized by this result -- a strain which breaks sublattice symmetry is required to induce a PMF, which in turn gives rise to a sublattice asymmetric DOS distribution.

\subsection{Finite size effects}
\begin{figure}
	\begin{center}
		\includegraphics[width= 0.95\columnwidth]{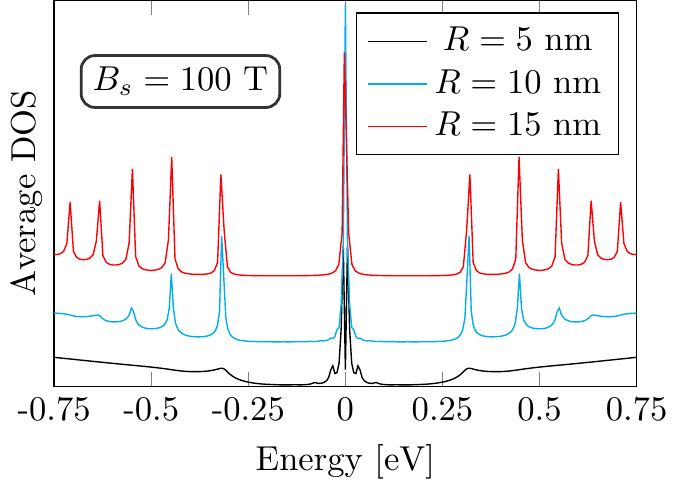}
		\caption[]{ \small Average density of states on sublattice B at the center of a triaxial strain corresponding to $B_s = 100$ T for radius $R=5,10$ and $15$ nm. 
			Each case has a smoothing region of width $\sigma=4$ nm and the curves are translated vertically with respect to each other. 
		} \label{fig:ldos_inplane_varR}
	\end{center}
\end{figure}
Finally, we investigate the influence of the size of the pseudomagnetic dot. 
\figref{ldos_inplane_varR} shows the LDOS on the B sublattice for a triaxial strain corresponding to $B_s=100$ T for three different radii $R=5,10$ and $15$ nm of the central region where the PMF is constant. 
In each case a smoothing is $\sigma=3$ nm is applied. 
We observe that the pLL peaks are broadened and almost disappear for the smallest dot size. 
Here the Landau quantization is washed out and the remaining states are expected to be unbound, as observed for the analogous real magnetic dot.\cite{RamezaniMasir2009}

To estimate when the dot becomes ``too small" to support proper Landau quantization we consider the corresponding magnetic length for the experienced PMF. 
At $B_s=100$ T we get a magnetic length of $l_B \sim 2.6$ nm from \eqref{lB}. 
For a radius significantly larger than $l_B$, we conclude from \figref{ldos_inplane_varR} that the pLLs are clearly formed. 
In contrast, the pLLs vanish when the strained region is comparable to or even slightly larger than $l_B$. 
This trend is observed for the $R=5$ nm case (black) in \figref{ldos_inplane_varR}. 
Instead, we notice here the formation of additional states around $E=0$ for small $R$. 
These are so called quasi bound states and are also observed for small real magnetic dots with varying field strengths \cite{RamezaniMasir2011}.
We conclude that the regions of constant PMF must be bigger than the corresponding magnetic length for pLLs to be clearly formed, even in the case of an idealized triaxial strain along the armchair direction.

\section{Conclusion} 
We employed a combination of analytical and numerical methods to analyse the behaviour of graphene systems subjected to PMFs. 
A mathematically transparent analysis, based on the effective low energy Dirac model, was used to explain the emergence of the sublattice polarization which has been predicted in a wide range of systems in the presence of PMFs \cite{Neek-Amal2013,Moldovan2013,Carrillo-Bastos2014,Settnes2015,Schneider2015}. 
This analysis was supported by a range of numerical calculations within a tight-binding model, which allowed us to confirm analytic predictions and explore finite size and rotation effects which are expected to be relevant in realistic systems.
Special attention was given to the zeroth pseudo Landau level, which was shown to be sublattice polarized.
This is in contrast to the valley polarization of individual sublattices expected for the case of a real magnetic field.
The occupied sublattice in the zeroth level for PMFs was furthermore shown to depend on the relative orientation of the strain and crystalline directions.
Numerical calculations performed in the absence of edges allowed us to confirm that this strong sublattice polarisation effect emerges entirely from pseudomagnetic considerations.
The strong rotational dependence of PMFs and their induced effects suggests that experimental systems with strains from multiple directions, such as circularly symmetric gas-inflated bubbles, will display inhomogeneous behavior that will be more difficult to characterize.  
In addition, we studied the breakdown of the analytic prediction of pLL for small dot sizes. 
Here we observe a broadening of the pseudo Landau peaks which disappear when the dot size is decreased towards the corresponding magnetic length. 
Understanding these limitations and effects are important in order to guide the exploitation of pseudomagnetic dots in strain engineering of graphene \cite{Jones2014,Zenan2014}. 

\textbf{Acknowledgements}
The Center for Nanostructured Graphene (CNG) is sponsored by the Danish Research Foundation, Project DNRF58.


\begin{thebibliography}{59}%
	\makeatletter
	\providecommand \@ifxundefined [1]{%
		\@ifx{#1\undefined}
	}%
	\providecommand \@ifnum [1]{%
		\ifnum #1\expandafter \@firstoftwo
		\else \expandafter \@secondoftwo
		\fi
	}%
	\providecommand \@ifx [1]{%
		\ifx #1\expandafter \@firstoftwo
		\else \expandafter \@secondoftwo
		\fi
	}%
	\providecommand \natexlab [1]{#1}%
	\providecommand \enquote  [1]{``#1''}%
	\providecommand \bibnamefont  [1]{#1}%
	\providecommand \bibfnamefont [1]{#1}%
	\providecommand \citenamefont [1]{#1}%
	\providecommand \href@noop [0]{\@secondoftwo}%
	\providecommand \href [0]{\begingroup \@sanitize@url \@href}%
	\providecommand \@href[1]{\@@startlink{#1}\@@href}%
	\providecommand \@@href[1]{\endgroup#1\@@endlink}%
	\providecommand \@sanitize@url [0]{\catcode `\\12\catcode `\$12\catcode
		`\&12\catcode `\#12\catcode `\^12\catcode `\_12\catcode `\%12\relax}%
	\providecommand \@@startlink[1]{}%
	\providecommand \@@endlink[0]{}%
	\providecommand \url  [0]{\begingroup\@sanitize@url \@url }%
	\providecommand \@url [1]{\endgroup\@href {#1}{\urlprefix }}%
	\providecommand \urlprefix  [0]{URL }%
	\providecommand \Eprint [0]{\href }%
	\providecommand \doibase [0]{http://dx.doi.org/}%
	\providecommand \selectlanguage [0]{\@gobble}%
	\providecommand \bibinfo  [0]{\@secondoftwo}%
	\providecommand \bibfield  [0]{\@secondoftwo}%
	\providecommand \translation [1]{[#1]}%
	\providecommand \BibitemOpen [0]{}%
	\providecommand \bibitemStop [0]{}%
	\providecommand \bibitemNoStop [0]{.\EOS\space}%
	\providecommand \EOS [0]{\spacefactor3000\relax}%
	\providecommand \BibitemShut  [1]{\csname bibitem#1\endcsname}%
	\let\auto@bib@innerbib\@empty
	\bibitem [{\citenamefont {Castro~Neto}\ \emph {et~al.}(2009)\citenamefont
		{Castro~Neto}, \citenamefont {Guinea}, \citenamefont {Peres}, \citenamefont
		{Novoselov},\ and\ \citenamefont {Geim}}]{CastroNeto2009}%
	\BibitemOpen
	\bibfield  {author} {\bibinfo {author} {\bibfnamefont {A.~H.}\ \bibnamefont
			{Castro~Neto}}, \bibinfo {author} {\bibfnamefont {F.}~\bibnamefont {Guinea}},
		\bibinfo {author} {\bibfnamefont {N.~M.~R.}\ \bibnamefont {Peres}}, \bibinfo
		{author} {\bibfnamefont {K.~S.}\ \bibnamefont {Novoselov}}, \ and\ \bibinfo
		{author} {\bibfnamefont {A.~K.}\ \bibnamefont {Geim}},\ }\href {\doibase
		10.1103/RevModPhys.81.109} {\bibfield  {journal} {\bibinfo  {journal} {Rev.
				Mod. Phys.}\ }\textbf {\bibinfo {volume} {81}},\ \bibinfo {pages} {109}
		(\bibinfo {year} {2009})}\BibitemShut {NoStop}%
	\bibitem [{\citenamefont {Vozmediano}\ \emph {et~al.}(2010)\citenamefont
		{Vozmediano}, \citenamefont {Katsnelson},\ and\ \citenamefont
		{Guinea}}]{Vozmediano2010}%
	\BibitemOpen
	\bibfield  {author} {\bibinfo {author} {\bibfnamefont {M.}~\bibnamefont
			{Vozmediano}}, \bibinfo {author} {\bibfnamefont {M.}~\bibnamefont
			{Katsnelson}}, \ and\ \bibinfo {author} {\bibfnamefont {F.}~\bibnamefont
			{Guinea}},\ }\href
	{http://www.sciencedirect.com/science/article/pii/S0370157310001729}
	{\bibfield  {journal} {\bibinfo  {journal} {Physics Reports}\ }\textbf
		{\bibinfo {volume} {496}},\ \bibinfo {pages} {109} (\bibinfo {year}
		{2010})}\BibitemShut {NoStop}%
	\bibitem [{\citenamefont {Guinea}\ \emph {et~al.}(2009)\citenamefont {Guinea},
		\citenamefont {Katsnelson},\ and\ \citenamefont {Geim}}]{Guinea2009}%
	\BibitemOpen
	\bibfield  {author} {\bibinfo {author} {\bibfnamefont {F.}~\bibnamefont
			{Guinea}}, \bibinfo {author} {\bibfnamefont {M.~I.}\ \bibnamefont
			{Katsnelson}}, \ and\ \bibinfo {author} {\bibfnamefont {A.~K.}\ \bibnamefont
			{Geim}},\ }\href {\doibase 10.1038/nphys1420} {\bibfield  {journal} {\bibinfo
			{journal} {Nature Physics}\ }\textbf {\bibinfo {volume} {6}},\ \bibinfo
		{pages} {30} (\bibinfo {year} {2009})}\BibitemShut {NoStop}%
	\bibitem [{\citenamefont {Neek-Amal}\ \emph {et~al.}(2013)\citenamefont
		{Neek-Amal}, \citenamefont {Covaci}, \citenamefont {Shakouri},\ and\
		\citenamefont {Peeters}}]{Neek-Amal2013}%
	\BibitemOpen
	\bibfield  {author} {\bibinfo {author} {\bibfnamefont {M.}~\bibnamefont
			{Neek-Amal}}, \bibinfo {author} {\bibfnamefont {L.}~\bibnamefont {Covaci}},
		\bibinfo {author} {\bibfnamefont {K.}~\bibnamefont {Shakouri}}, \ and\
		\bibinfo {author} {\bibfnamefont {F.~M.}\ \bibnamefont {Peeters}},\ }\href
	{\doibase 10.1103/PhysRevB.88.115428} {\bibfield  {journal} {\bibinfo
			{journal} {Physical Review B}\ }\textbf {\bibinfo {volume} {88}},\ \bibinfo
		{pages} {115428} (\bibinfo {year} {2013})}\BibitemShut {NoStop}%
	\bibitem [{\citenamefont {Guinea}\ \emph {et~al.}(2010)\citenamefont {Guinea},
		\citenamefont {Geim}, \citenamefont {Katsnelson},\ and\ \citenamefont
		{Novoselov}}]{Guinea2010}%
	\BibitemOpen
	\bibfield  {author} {\bibinfo {author} {\bibfnamefont {F.}~\bibnamefont
			{Guinea}}, \bibinfo {author} {\bibfnamefont {A.~K.}\ \bibnamefont {Geim}},
		\bibinfo {author} {\bibfnamefont {M.~I.}\ \bibnamefont {Katsnelson}}, \ and\
		\bibinfo {author} {\bibfnamefont {K.~S.}\ \bibnamefont {Novoselov}},\ }\href
	{\doibase 10.1103/PhysRevB.81.035408} {\bibfield  {journal} {\bibinfo
			{journal} {Phys. Rev. B}\ }\textbf {\bibinfo {volume} {81}},\ \bibinfo
		{pages} {035408} (\bibinfo {year} {2010})}\BibitemShut {NoStop}%
	\bibitem [{\citenamefont {Pereira}\ \emph {et~al.}(2009)\citenamefont
		{Pereira}, \citenamefont {{Castro Neto}},\ and\ \citenamefont
		{Peres}}]{Pereira2009}%
	\BibitemOpen
	\bibfield  {author} {\bibinfo {author} {\bibfnamefont {V.~M.}\ \bibnamefont
			{Pereira}}, \bibinfo {author} {\bibfnamefont {A.~H.}\ \bibnamefont {{Castro
					Neto}}}, \ and\ \bibinfo {author} {\bibfnamefont {N.~M.~R.}\ \bibnamefont
			{Peres}},\ }\href {\doibase 10.1103/PhysRevB.80.045401} {\bibfield  {journal}
		{\bibinfo  {journal} {Physical Review B}\ }\textbf {\bibinfo {volume} {80}},\
		\bibinfo {pages} {045401} (\bibinfo {year} {2009})}\BibitemShut {NoStop}%
	\bibitem [{\citenamefont {Guinea}(2012)}]{Guinea2012}%
	\BibitemOpen
	\bibfield  {author} {\bibinfo {author} {\bibfnamefont {F.}~\bibnamefont
			{Guinea}},\ }\href {\doibase http://dx.doi.org/10.1016/j.ssc.2012.04.019}
	{\bibfield  {journal} {\bibinfo  {journal} {Solid State Communications}\
		}\textbf {\bibinfo {volume} {152}},\ \bibinfo {pages} {1437 } (\bibinfo
		{year} {2012})}\BibitemShut {NoStop}%
	\bibitem [{\citenamefont {Jones}\ and\ \citenamefont
		{Pereira}(2014)}]{Jones2014}%
	\BibitemOpen
	\bibfield  {author} {\bibinfo {author} {\bibfnamefont {G.~W.}\ \bibnamefont
			{Jones}}\ and\ \bibinfo {author} {\bibfnamefont {V.~M.}\ \bibnamefont
			{Pereira}},\ }\href {\doibase 10.1088/1367-2630/16/9/093044} {\bibfield
		{journal} {\bibinfo  {journal} {New Journal of Physics}\ }\textbf {\bibinfo
			{volume} {16}},\ \bibinfo {pages} {093044} (\bibinfo {year}
		{2014})}\BibitemShut {NoStop}%
	\bibitem [{\citenamefont {Chaves}\ \emph {et~al.}(2010)\citenamefont {Chaves},
		\citenamefont {Covaci}, \citenamefont {Rakhimov}, \citenamefont {Farias},\
		and\ \citenamefont {Peeters}}]{Chaves2010}%
	\BibitemOpen
	\bibfield  {author} {\bibinfo {author} {\bibfnamefont {A.}~\bibnamefont
			{Chaves}}, \bibinfo {author} {\bibfnamefont {L.}~\bibnamefont {Covaci}},
		\bibinfo {author} {\bibfnamefont {K.~Y.}\ \bibnamefont {Rakhimov}}, \bibinfo
		{author} {\bibfnamefont {G.~A.}\ \bibnamefont {Farias}}, \ and\ \bibinfo
		{author} {\bibfnamefont {F.~M.}\ \bibnamefont {Peeters}},\ }\href {\doibase
		10.1103/PhysRevB.82.205430} {\bibfield  {journal} {\bibinfo  {journal} {Phys.
				Rev. B}\ }\textbf {\bibinfo {volume} {82}},\ \bibinfo {pages} {205430}
		(\bibinfo {year} {2010})}\BibitemShut {NoStop}%
	\bibitem [{\citenamefont {Low}\ and\ \citenamefont {Guinea}(2010)}]{Low2010}%
	\BibitemOpen
	\bibfield  {author} {\bibinfo {author} {\bibfnamefont {T.}~\bibnamefont
			{Low}}\ and\ \bibinfo {author} {\bibfnamefont {F.}~\bibnamefont {Guinea}},\
	}\href {\doibase 10.1021/nl1018063} {\bibfield  {journal} {\bibinfo
		{journal} {Nano letters}\ }\textbf {\bibinfo {volume} {10}},\ \bibinfo
	{pages} {3551} (\bibinfo {year} {2010})}\BibitemShut {NoStop}%
\bibitem [{\citenamefont {Wu}\ \emph {et~al.}(2011)\citenamefont {Wu},
	\citenamefont {Zhai}, \citenamefont {Peeters}, \citenamefont {Xu},\ and\
	\citenamefont {Chang}}]{Zhenhua2011}%
\BibitemOpen
\bibfield  {author} {\bibinfo {author} {\bibfnamefont {Z.}~\bibnamefont
		{Wu}}, \bibinfo {author} {\bibfnamefont {F.}~\bibnamefont {Zhai}}, \bibinfo
	{author} {\bibfnamefont {F.~M.}\ \bibnamefont {Peeters}}, \bibinfo {author}
	{\bibfnamefont {H.~Q.}\ \bibnamefont {Xu}}, \ and\ \bibinfo {author}
	{\bibfnamefont {K.}~\bibnamefont {Chang}},\ }\href {\doibase
	10.1103/PhysRevLett.106.176802} {\bibfield  {journal} {\bibinfo  {journal}
		{Phys. Rev. Lett.}\ }\textbf {\bibinfo {volume} {106}},\ \bibinfo {pages}
	{176802} (\bibinfo {year} {2011})}\BibitemShut {NoStop}%
\bibitem [{\citenamefont {Kim}\ \emph {et~al.}(2011)\citenamefont {Kim},
	\citenamefont {Blanter},\ and\ \citenamefont {Ahn}}]{Kim2011}%
\BibitemOpen
\bibfield  {author} {\bibinfo {author} {\bibfnamefont {K.-J.}\ \bibnamefont
		{Kim}}, \bibinfo {author} {\bibfnamefont {Y.~M.}\ \bibnamefont {Blanter}}, \
	and\ \bibinfo {author} {\bibfnamefont {K.-H.}\ \bibnamefont {Ahn}},\ }\href
{\doibase 10.1103/PhysRevB.84.081401} {\bibfield  {journal} {\bibinfo
		{journal} {Physical Review B}\ }\textbf {\bibinfo {volume} {84}},\ \bibinfo
	{pages} {081401} (\bibinfo {year} {2011})}\BibitemShut {NoStop}%
\bibitem [{\citenamefont {Qi}\ \emph {et~al.}(2013)\citenamefont {Qi},
	\citenamefont {Bahamon}, \citenamefont {Pereira}, \citenamefont {Park},
	\citenamefont {Campbell},\ and\ \citenamefont {{Castro Neto}}}]{Qi2013}%
\BibitemOpen
\bibfield  {author} {\bibinfo {author} {\bibfnamefont {Z.}~\bibnamefont
		{Qi}}, \bibinfo {author} {\bibfnamefont {D.~A.}\ \bibnamefont {Bahamon}},
	\bibinfo {author} {\bibfnamefont {V.~M.}\ \bibnamefont {Pereira}}, \bibinfo
	{author} {\bibfnamefont {H.~S.}\ \bibnamefont {Park}}, \bibinfo {author}
	{\bibfnamefont {D.~K.}\ \bibnamefont {Campbell}}, \ and\ \bibinfo {author}
	{\bibfnamefont {A.~H.}\ \bibnamefont {{Castro Neto}}},\ }\href {\doibase
	10.1021/nl400872q} {\bibfield  {journal} {\bibinfo  {journal} {Nano letters}\
	}\textbf {\bibinfo {volume} {13}},\ \bibinfo {pages} {2692} (\bibinfo {year}
	{2013})}\BibitemShut {NoStop}%
\bibitem [{\citenamefont {Ni}\ \emph {et~al.}(2009)\citenamefont {Ni},
	\citenamefont {Yu}, \citenamefont {Lu}, \citenamefont {Wang}, \citenamefont
	{Feng},\ and\ \citenamefont {Shen}}]{Ni2009}%
\BibitemOpen
\bibfield  {author} {\bibinfo {author} {\bibfnamefont {Z.~H.}\ \bibnamefont
		{Ni}}, \bibinfo {author} {\bibfnamefont {T.}~\bibnamefont {Yu}}, \bibinfo
	{author} {\bibfnamefont {Y.~H.}\ \bibnamefont {Lu}}, \bibinfo {author}
	{\bibfnamefont {Y.~Y.}\ \bibnamefont {Wang}}, \bibinfo {author}
	{\bibfnamefont {Y.~P.}\ \bibnamefont {Feng}}, \ and\ \bibinfo {author}
	{\bibfnamefont {Z.~X.}\ \bibnamefont {Shen}},\ }\href {\doibase
	10.1021/nn8008323} {\bibfield  {journal} {\bibinfo  {journal} {ACS Nano}\
	}\textbf {\bibinfo {volume} {3}},\ \bibinfo {pages} {483} (\bibinfo {year}
	{2009})}\BibitemShut {NoStop}%
\bibitem [{\citenamefont {Low}\ \emph {et~al.}(2011)\citenamefont {Low},
	\citenamefont {Guinea},\ and\ \citenamefont {Katsnelson}}]{Low2011}%
\BibitemOpen
\bibfield  {author} {\bibinfo {author} {\bibfnamefont {T.}~\bibnamefont
		{Low}}, \bibinfo {author} {\bibfnamefont {F.}~\bibnamefont {Guinea}}, \ and\
	\bibinfo {author} {\bibfnamefont {M.~I.}\ \bibnamefont {Katsnelson}},\ }\href
{\doibase 10.1103/PhysRevB.83.195436} {\bibfield  {journal} {\bibinfo
		{journal} {Physical Review B}\ }\textbf {\bibinfo {volume} {83}},\ \bibinfo
	{pages} {195436} (\bibinfo {year} {2011})}\BibitemShut {NoStop}%
\bibitem [{\citenamefont {Levy}\ \emph {et~al.}(2010)\citenamefont {Levy},
	\citenamefont {Burke}, \citenamefont {Meaker}, \citenamefont {Panlasigui},
	\citenamefont {Zettl}, \citenamefont {Guinea}, \citenamefont {{Castro
			Neto}},\ and\ \citenamefont {Crommie}}]{Levy2010}%
\BibitemOpen
\bibfield  {author} {\bibinfo {author} {\bibfnamefont {N.}~\bibnamefont
		{Levy}}, \bibinfo {author} {\bibfnamefont {S.~A.}\ \bibnamefont {Burke}},
	\bibinfo {author} {\bibfnamefont {K.~L.}\ \bibnamefont {Meaker}}, \bibinfo
	{author} {\bibfnamefont {M.}~\bibnamefont {Panlasigui}}, \bibinfo {author}
	{\bibfnamefont {A.}~\bibnamefont {Zettl}}, \bibinfo {author} {\bibfnamefont
		{F.}~\bibnamefont {Guinea}}, \bibinfo {author} {\bibfnamefont {A.~H.}\
		\bibnamefont {{Castro Neto}}}, \ and\ \bibinfo {author} {\bibfnamefont
		{M.~F.}\ \bibnamefont {Crommie}},\ }\href {\doibase 10.1126/science.1191700}
{\bibfield  {journal} {\bibinfo  {journal} {Science}\ }\textbf {\bibinfo
		{volume} {329}},\ \bibinfo {pages} {544} (\bibinfo {year}
	{2010})}\BibitemShut {NoStop}%
\bibitem [{\citenamefont {Lu}\ \emph {et~al.}(2012)\citenamefont {Lu},
	\citenamefont {{Castro Neto}},\ and\ \citenamefont {Loh}}]{Lu2012}%
\BibitemOpen
\bibfield  {author} {\bibinfo {author} {\bibfnamefont {J.}~\bibnamefont
		{Lu}}, \bibinfo {author} {\bibfnamefont {A.~H.}\ \bibnamefont {{Castro
				Neto}}}, \ and\ \bibinfo {author} {\bibfnamefont {K.~P.}\ \bibnamefont
		{Loh}},\ }\href {\doibase 10.1038/ncomms1818} {\bibfield  {journal} {\bibinfo
		{journal} {Nature communications}\ }\textbf {\bibinfo {volume} {3}},\
	\bibinfo {pages} {823} (\bibinfo {year} {2012})}\BibitemShut {NoStop}%
\bibitem [{\citenamefont {Klimov}\ \emph {et~al.}(2012)\citenamefont {Klimov},
	\citenamefont {Jung}, \citenamefont {Zhu}, \citenamefont {Li}, \citenamefont
	{Wright}, \citenamefont {Solares}, \citenamefont {Newell}, \citenamefont
	{Zhitenev},\ and\ \citenamefont {Stroscio}}]{Klimov2012}%
\BibitemOpen
\bibfield  {author} {\bibinfo {author} {\bibfnamefont {N.~N.}\ \bibnamefont
		{Klimov}}, \bibinfo {author} {\bibfnamefont {S.}~\bibnamefont {Jung}},
	\bibinfo {author} {\bibfnamefont {S.}~\bibnamefont {Zhu}}, \bibinfo {author}
	{\bibfnamefont {T.}~\bibnamefont {Li}}, \bibinfo {author} {\bibfnamefont
		{C.~A.}\ \bibnamefont {Wright}}, \bibinfo {author} {\bibfnamefont {S.~D.}\
		\bibnamefont {Solares}}, \bibinfo {author} {\bibfnamefont {D.~B.}\
		\bibnamefont {Newell}}, \bibinfo {author} {\bibfnamefont {N.~B.}\
		\bibnamefont {Zhitenev}}, \ and\ \bibinfo {author} {\bibfnamefont {J.~A.}\
		\bibnamefont {Stroscio}},\ }\href {\doibase 10.1126/science.1220335}
{\bibfield  {journal} {\bibinfo  {journal} {Science}\ }\textbf {\bibinfo
		{volume} {336}},\ \bibinfo {pages} {1557} (\bibinfo {year}
	{2012})}\BibitemShut {NoStop}%
\bibitem [{\citenamefont {Bunch}\ \emph {et~al.}(2008)\citenamefont {Bunch},
	\citenamefont {Verbridge}, \citenamefont {Alden}, \citenamefont {van~der
		Zande}, \citenamefont {Parpia}, \citenamefont {Craighead},\ and\
	\citenamefont {McEuen}}]{Bunch2008}%
\BibitemOpen
\bibfield  {author} {\bibinfo {author} {\bibfnamefont {J.~S.}\ \bibnamefont
		{Bunch}}, \bibinfo {author} {\bibfnamefont {S.~S.}\ \bibnamefont
		{Verbridge}}, \bibinfo {author} {\bibfnamefont {J.~S.}\ \bibnamefont
		{Alden}}, \bibinfo {author} {\bibfnamefont {A.~M.}\ \bibnamefont {van~der
			Zande}}, \bibinfo {author} {\bibfnamefont {J.~M.}\ \bibnamefont {Parpia}},
	\bibinfo {author} {\bibfnamefont {H.~G.}\ \bibnamefont {Craighead}}, \ and\
	\bibinfo {author} {\bibfnamefont {P.~L.}\ \bibnamefont {McEuen}},\ }\href
{\doibase 10.1021/nl801457b} {\bibfield  {journal} {\bibinfo  {journal} {Nano
			Letters}\ }\textbf {\bibinfo {volume} {8}},\ \bibinfo {pages} {2458}
	(\bibinfo {year} {2008})}\BibitemShut {NoStop}%
\bibitem [{\citenamefont {Georgiou}\ \emph {et~al.}(2011)\citenamefont
	{Georgiou}, \citenamefont {Britnell}, \citenamefont {Blake}, \citenamefont
	{Gorbachev}, \citenamefont {Gholinia}, \citenamefont {Geim}, \citenamefont
	{Casiraghi},\ and\ \citenamefont {Novoselov}}]{Georgiou2011}%
\BibitemOpen
\bibfield  {author} {\bibinfo {author} {\bibfnamefont {T.}~\bibnamefont
		{Georgiou}}, \bibinfo {author} {\bibfnamefont {L.}~\bibnamefont {Britnell}},
	\bibinfo {author} {\bibfnamefont {P.}~\bibnamefont {Blake}}, \bibinfo
	{author} {\bibfnamefont {R.~V.}\ \bibnamefont {Gorbachev}}, \bibinfo {author}
	{\bibfnamefont {A.}~\bibnamefont {Gholinia}}, \bibinfo {author}
	{\bibfnamefont {A.~K.}\ \bibnamefont {Geim}}, \bibinfo {author}
	{\bibfnamefont {C.}~\bibnamefont {Casiraghi}}, \ and\ \bibinfo {author}
	{\bibfnamefont {K.~S.}\ \bibnamefont {Novoselov}},\ }\href {\doibase
	http://dx.doi.org/10.1063/1.3631632} {\bibfield  {journal} {\bibinfo
		{journal} {Applied Physics Letters}\ }\textbf {\bibinfo {volume} {99}},\
	\bibinfo {eid} {093103} (\bibinfo {year} {2011})}\BibitemShut {NoStop}%
\bibitem [{\citenamefont {Qi}\ \emph {et~al.}(2014)\citenamefont {Qi},
	\citenamefont {Kitt}, \citenamefont {Park}, \citenamefont {Pereira},
	\citenamefont {Campbell},\ and\ \citenamefont {{Castro Neto}}}]{Zenan2014}%
\BibitemOpen
\bibfield  {author} {\bibinfo {author} {\bibfnamefont {Z.}~\bibnamefont
		{Qi}}, \bibinfo {author} {\bibfnamefont {A.~L.}\ \bibnamefont {Kitt}},
	\bibinfo {author} {\bibfnamefont {H.~S.}\ \bibnamefont {Park}}, \bibinfo
	{author} {\bibfnamefont {V.~M.}\ \bibnamefont {Pereira}}, \bibinfo {author}
	{\bibfnamefont {D.~K.}\ \bibnamefont {Campbell}}, \ and\ \bibinfo {author}
	{\bibfnamefont {A.~H.}\ \bibnamefont {{Castro Neto}}},\ }\href {\doibase
	10.1103/PhysRevB.90.125419} {\bibfield  {journal} {\bibinfo  {journal} {Phys.
			Rev. B}\ }\textbf {\bibinfo {volume} {90}},\ \bibinfo {pages} {125419}
	(\bibinfo {year} {2014})}\BibitemShut {NoStop}%
\bibitem [{\citenamefont {Settnes}\ \emph
	{et~al.}(2015{\natexlab{a}})\citenamefont {Settnes}, \citenamefont {Power},
	\citenamefont {Lin}, \citenamefont {Petersen},\ and\ \citenamefont
	{Jauho}}]{Settnes2015}%
\BibitemOpen
\bibfield  {author} {\bibinfo {author} {\bibfnamefont {M.}~\bibnamefont
		{Settnes}}, \bibinfo {author} {\bibfnamefont {S.~R.}\ \bibnamefont {Power}},
	\bibinfo {author} {\bibfnamefont {J.}~\bibnamefont {Lin}}, \bibinfo {author}
	{\bibfnamefont {D.~H.}\ \bibnamefont {Petersen}}, \ and\ \bibinfo {author}
	{\bibfnamefont {A.-P.}\ \bibnamefont {Jauho}},\ }\href {\doibase
	10.1103/PhysRevB.91.125408} {\bibfield  {journal} {\bibinfo  {journal}
		{Physical Review B}\ }\textbf {\bibinfo {volume} {91}},\ \bibinfo {pages}
	{125408} (\bibinfo {year} {2015}{\natexlab{a}})}\BibitemShut {NoStop}%
\bibitem [{\citenamefont {Bahamon}\ \emph {et~al.}(2015)\citenamefont
	{Bahamon}, \citenamefont {Qi}, \citenamefont {Park}, \citenamefont
	{Pereira},\ and\ \citenamefont {Campbell}}]{Bahamon2015}%
\BibitemOpen
\bibfield  {author} {\bibinfo {author} {\bibfnamefont {D.~A.}\ \bibnamefont
		{Bahamon}}, \bibinfo {author} {\bibfnamefont {Z.}~\bibnamefont {Qi}},
	\bibinfo {author} {\bibfnamefont {H.~S.}\ \bibnamefont {Park}}, \bibinfo
	{author} {\bibfnamefont {V.~M.}\ \bibnamefont {Pereira}}, \ and\ \bibinfo
	{author} {\bibfnamefont {D.~K.}\ \bibnamefont {Campbell}},\ }\href {\doibase
	10.1039/C5NR03393D} {\bibfield  {journal} {\bibinfo  {journal} {Nanoscale}\
		,\ } (\bibinfo {year} {2015})}\BibitemShut {NoStop}%
\bibitem [{\citenamefont {Tomori}\ \emph {et~al.}(2011)\citenamefont {Tomori},
	\citenamefont {Kanda}, \citenamefont {Goto}, \citenamefont {Ootuka},
	\citenamefont {Tsukagoshi}, \citenamefont {Moriyama}, \citenamefont
	{Watanabe},\ and\ \citenamefont {Tsuya}}]{Tomori2011}%
\BibitemOpen
\bibfield  {author} {\bibinfo {author} {\bibfnamefont {H.}~\bibnamefont
		{Tomori}}, \bibinfo {author} {\bibfnamefont {A.}~\bibnamefont {Kanda}},
	\bibinfo {author} {\bibfnamefont {H.}~\bibnamefont {Goto}}, \bibinfo {author}
	{\bibfnamefont {Y.}~\bibnamefont {Ootuka}}, \bibinfo {author} {\bibfnamefont
		{K.}~\bibnamefont {Tsukagoshi}}, \bibinfo {author} {\bibfnamefont
		{S.}~\bibnamefont {Moriyama}}, \bibinfo {author} {\bibfnamefont
		{E.}~\bibnamefont {Watanabe}}, \ and\ \bibinfo {author} {\bibfnamefont
		{D.}~\bibnamefont {Tsuya}},\ }\href
{http://stacks.iop.org/1882-0786/4/i=7/a=075102} {\bibfield  {journal}
	{\bibinfo  {journal} {Applied Physics Express}\ }\textbf {\bibinfo {volume}
		{4}},\ \bibinfo {pages} {075102} (\bibinfo {year} {2011})}\BibitemShut
{NoStop}%
\bibitem [{\citenamefont {Reserbat-Plantey}\ \emph {et~al.}(2014)\citenamefont
	{Reserbat-Plantey}, \citenamefont {Kalita}, \citenamefont {Han},
	\citenamefont {Ferlazzo}, \citenamefont {Autier-Laurent}, \citenamefont
	{Komatsu}, \citenamefont {Li}, \citenamefont {Weil}, \citenamefont {Ralko},
	\citenamefont {Marty}, \citenamefont {Guéron}, \citenamefont {Bendiab},
	\citenamefont {Bouchiat},\ and\ \citenamefont
	{Bouchiat}}]{Reserbat-Plantey2014}%
\BibitemOpen
\bibfield  {author} {\bibinfo {author} {\bibfnamefont {A.}~\bibnamefont
		{Reserbat-Plantey}}, \bibinfo {author} {\bibfnamefont {D.}~\bibnamefont
		{Kalita}}, \bibinfo {author} {\bibfnamefont {Z.}~\bibnamefont {Han}},
	\bibinfo {author} {\bibfnamefont {L.}~\bibnamefont {Ferlazzo}}, \bibinfo
	{author} {\bibfnamefont {S.}~\bibnamefont {Autier-Laurent}}, \bibinfo
	{author} {\bibfnamefont {K.}~\bibnamefont {Komatsu}}, \bibinfo {author}
	{\bibfnamefont {C.}~\bibnamefont {Li}}, \bibinfo {author} {\bibfnamefont
		{R.}~\bibnamefont {Weil}}, \bibinfo {author} {\bibfnamefont {A.}~\bibnamefont
		{Ralko}}, \bibinfo {author} {\bibfnamefont {L.}~\bibnamefont {Marty}},
	\bibinfo {author} {\bibfnamefont {S.}~\bibnamefont {Guéron}}, \bibinfo
	{author} {\bibfnamefont {N.}~\bibnamefont {Bendiab}}, \bibinfo {author}
	{\bibfnamefont {H.}~\bibnamefont {Bouchiat}}, \ and\ \bibinfo {author}
	{\bibfnamefont {V.}~\bibnamefont {Bouchiat}},\ }\href {\doibase
	10.1021/nl5016552} {\bibfield  {journal} {\bibinfo  {journal} {Nano Letters}\
	}\textbf {\bibinfo {volume} {14}},\ \bibinfo {pages} {5044} (\bibinfo {year}
	{2014})}\BibitemShut {NoStop}%
\bibitem [{\citenamefont {Gill}\ \emph {et~al.}(2015)\citenamefont {Gill},
	\citenamefont {Hinnefeld}, \citenamefont {Zhu}, \citenamefont {Swanson},
	\citenamefont {Li},\ and\ \citenamefont {Mason}}]{Gill2015}%
\BibitemOpen
\bibfield  {author} {\bibinfo {author} {\bibfnamefont {S.~T.}\ \bibnamefont
		{Gill}}, \bibinfo {author} {\bibfnamefont {J.~H.}\ \bibnamefont {Hinnefeld}},
	\bibinfo {author} {\bibfnamefont {S.}~\bibnamefont {Zhu}}, \bibinfo {author}
	{\bibfnamefont {W.~J.}\ \bibnamefont {Swanson}}, \bibinfo {author}
	{\bibfnamefont {T.}~\bibnamefont {Li}}, \ and\ \bibinfo {author}
	{\bibfnamefont {N.}~\bibnamefont {Mason}},\ }\href {\doibase
	10.1021/acsnano.5b00335} {\bibfield  {journal} {\bibinfo  {journal} {ACS
			Nano}\ }\textbf {\bibinfo {volume} {9}},\ \bibinfo {pages} {5799} (\bibinfo
	{year} {2015})}\BibitemShut {NoStop}%
\bibitem [{\citenamefont {Scharfenberg}\ \emph {et~al.}(2011)\citenamefont
	{Scharfenberg}, \citenamefont {Rocklin}, \citenamefont {Chialvo},
	\citenamefont {Weaver}, \citenamefont {Goldbart},\ and\ \citenamefont
	{Mason}}]{Scharfenberg2011}%
\BibitemOpen
\bibfield  {author} {\bibinfo {author} {\bibfnamefont {S.}~\bibnamefont
		{Scharfenberg}}, \bibinfo {author} {\bibfnamefont {D.~Z.}\ \bibnamefont
		{Rocklin}}, \bibinfo {author} {\bibfnamefont {C.}~\bibnamefont {Chialvo}},
	\bibinfo {author} {\bibfnamefont {R.~L.}\ \bibnamefont {Weaver}}, \bibinfo
	{author} {\bibfnamefont {P.~M.}\ \bibnamefont {Goldbart}}, \ and\ \bibinfo
	{author} {\bibfnamefont {N.}~\bibnamefont {Mason}},\ }\href {\doibase
	http://dx.doi.org/10.1063/1.3553228} {\bibfield  {journal} {\bibinfo
		{journal} {Applied Physics Letters}\ }\textbf {\bibinfo {volume} {98}},\
	\bibinfo {eid} {091908} (\bibinfo {year} {2011})}\BibitemShut {NoStop}%
\bibitem [{\citenamefont {Zang}\ \emph {et~al.}(2013)\citenamefont {Zang},
	\citenamefont {Ryu}, \citenamefont {Pugno}, \citenamefont {Wang},
	\citenamefont {Tu},\ and\ \citenamefont {Buehler}}]{Zang2013}%
\BibitemOpen
\bibfield  {author} {\bibinfo {author} {\bibfnamefont {J.}~\bibnamefont
		{Zang}}, \bibinfo {author} {\bibfnamefont {S.}~\bibnamefont {Ryu}}, \bibinfo
	{author} {\bibfnamefont {N.}~\bibnamefont {Pugno}}, \bibinfo {author}
	{\bibfnamefont {Q.}~\bibnamefont {Wang}}, \bibinfo {author} {\bibfnamefont
		{Q.}~\bibnamefont {Tu}}, \ and\ \bibinfo {author} {\bibfnamefont {M.~J.}\
		\bibnamefont {Buehler}},\ }\href {\doibase 10.1038/nmat3542} {\bibfield
	{journal} {\bibinfo  {journal} {Nature Materials}\ }\textbf {\bibinfo
		{volume} {12}},\ \bibinfo {pages} {321} (\bibinfo {year} {2013})}\BibitemShut
{NoStop}%
\bibitem [{\citenamefont {Bao}\ \emph {et~al.}(2009)\citenamefont {Bao},
	\citenamefont {Miao}, \citenamefont {Chen}, \citenamefont {Zhang},
	\citenamefont {Jang},\ and\ \citenamefont {Dames}}]{Bao2009}%
\BibitemOpen
\bibfield  {author} {\bibinfo {author} {\bibfnamefont {W.}~\bibnamefont
		{Bao}}, \bibinfo {author} {\bibfnamefont {F.}~\bibnamefont {Miao}}, \bibinfo
	{author} {\bibfnamefont {Z.}~\bibnamefont {Chen}}, \bibinfo {author}
	{\bibfnamefont {H.}~\bibnamefont {Zhang}}, \bibinfo {author} {\bibfnamefont
		{W.}~\bibnamefont {Jang}}, \ and\ \bibinfo {author} {\bibfnamefont
		{C.}~\bibnamefont {Dames}},\ }\href {\doibase 10.1038/nnano.2009.191}
{\bibfield  {journal} {\bibinfo  {journal} {Nature Nanotechnology}\ }\textbf
	{\bibinfo {volume} {4}},\ \bibinfo {pages} {562} (\bibinfo {year}
	{2009})}\BibitemShut {NoStop}%
\bibitem [{\citenamefont {Meng}\ \emph {et~al.}(2013)\citenamefont {Meng},
	\citenamefont {He}, \citenamefont {Zheng}, \citenamefont {Liu}, \citenamefont
	{Yan}, \citenamefont {Yan}, \citenamefont {Chu}, \citenamefont {Bai},
	\citenamefont {Dou}, \citenamefont {Zhang}, \citenamefont {Liu},
	\citenamefont {Nie},\ and\ \citenamefont {He}}]{Meng2013}%
\BibitemOpen
\bibfield  {author} {\bibinfo {author} {\bibfnamefont {L.}~\bibnamefont
		{Meng}}, \bibinfo {author} {\bibfnamefont {W.-Y.}\ \bibnamefont {He}},
	\bibinfo {author} {\bibfnamefont {H.}~\bibnamefont {Zheng}}, \bibinfo
	{author} {\bibfnamefont {M.}~\bibnamefont {Liu}}, \bibinfo {author}
	{\bibfnamefont {H.}~\bibnamefont {Yan}}, \bibinfo {author} {\bibfnamefont
		{W.}~\bibnamefont {Yan}}, \bibinfo {author} {\bibfnamefont {Z.-D.}\
		\bibnamefont {Chu}}, \bibinfo {author} {\bibfnamefont {K.}~\bibnamefont
		{Bai}}, \bibinfo {author} {\bibfnamefont {R.-F.}\ \bibnamefont {Dou}},
	\bibinfo {author} {\bibfnamefont {Y.}~\bibnamefont {Zhang}}, \bibinfo
	{author} {\bibfnamefont {Z.}~\bibnamefont {Liu}}, \bibinfo {author}
	{\bibfnamefont {J.-C.}\ \bibnamefont {Nie}}, \ and\ \bibinfo {author}
	{\bibfnamefont {L.}~\bibnamefont {He}},\ }\href {\doibase
	10.1103/PhysRevB.87.205405} {\bibfield  {journal} {\bibinfo  {journal} {Phys.
			Rev. B}\ }\textbf {\bibinfo {volume} {87}},\ \bibinfo {pages} {205405}
	(\bibinfo {year} {2013})}\BibitemShut {NoStop}%
\bibitem [{\citenamefont {Neek-Amal}\ \emph {et~al.}(2012)\citenamefont
	{Neek-Amal}, \citenamefont {Covaci},\ and\ \citenamefont
	{Peeters}}]{Neek-Amal2012c}%
\BibitemOpen
\bibfield  {author} {\bibinfo {author} {\bibfnamefont {M.}~\bibnamefont
		{Neek-Amal}}, \bibinfo {author} {\bibfnamefont {L.}~\bibnamefont {Covaci}}, \
	and\ \bibinfo {author} {\bibfnamefont {F.~M.}\ \bibnamefont {Peeters}},\
}\href {\doibase 10.1103/PhysRevB.86.041405} {\bibfield  {journal} {\bibinfo
	{journal} {Physical Review B}\ }\textbf {\bibinfo {volume} {86}},\ \bibinfo
{pages} {041405} (\bibinfo {year} {2012})}\BibitemShut {NoStop}%
\bibitem [{\citenamefont {Sloan}\ \emph {et~al.}(2013)\citenamefont {Sloan},
	\citenamefont {Sanjuan}, \citenamefont {Wang}, \citenamefont {Horvath},\ and\
	\citenamefont {Barraza-Lopez}}]{Sloan2013}%
\BibitemOpen
\bibfield  {author} {\bibinfo {author} {\bibfnamefont {J.~V.}\ \bibnamefont
		{Sloan}}, \bibinfo {author} {\bibfnamefont {A.~A.}\ \bibnamefont {Sanjuan}},
	\bibinfo {author} {\bibfnamefont {Z.}~\bibnamefont {Wang}}, \bibinfo {author}
	{\bibfnamefont {C.}~\bibnamefont {Horvath}}, \ and\ \bibinfo {author}
	{\bibfnamefont {S.}~\bibnamefont {Barraza-Lopez}},\ }\href {\doibase
	10.1103/PhysRevB.87.155436} {\bibfield  {journal} {\bibinfo  {journal} {Phys.
			Rev. B}\ }\textbf {\bibinfo {volume} {87}},\ \bibinfo {pages} {155436}
	(\bibinfo {year} {2013})}\BibitemShut {NoStop}%
\bibitem [{\citenamefont {Shioya}\ \emph {et~al.}(2014)\citenamefont {Shioya},
	\citenamefont {Craciun}, \citenamefont {Russo}, \citenamefont {Yamamoto},\
	and\ \citenamefont {Tarucha}}]{Shioya2014}%
\BibitemOpen
\bibfield  {author} {\bibinfo {author} {\bibfnamefont {H.}~\bibnamefont
		{Shioya}}, \bibinfo {author} {\bibfnamefont {M.~F.}\ \bibnamefont {Craciun}},
	\bibinfo {author} {\bibfnamefont {S.}~\bibnamefont {Russo}}, \bibinfo
	{author} {\bibfnamefont {M.}~\bibnamefont {Yamamoto}}, \ and\ \bibinfo
	{author} {\bibfnamefont {S.}~\bibnamefont {Tarucha}},\ }\href {\doibase
	10.1021/nl403679f} {\bibfield  {journal} {\bibinfo  {journal} {Nano Letters}\
	}\textbf {\bibinfo {volume} {14}},\ \bibinfo {pages} {1158} (\bibinfo {year}
	{2014})}\BibitemShut {NoStop}%
\bibitem [{\citenamefont {Neek-Amal}\ and\ \citenamefont
	{Peeters}(2012{\natexlab{a}})}]{Neek-Amal2012a}%
\BibitemOpen
\bibfield  {author} {\bibinfo {author} {\bibfnamefont {M.}~\bibnamefont
		{Neek-Amal}}\ and\ \bibinfo {author} {\bibfnamefont {F.~M.}\ \bibnamefont
		{Peeters}},\ }\href {\doibase 10.1103/PhysRevB.85.195445} {\bibfield
	{journal} {\bibinfo  {journal} {Phys. Rev. B}\ }\textbf {\bibinfo {volume}
		{85}},\ \bibinfo {pages} {195445} (\bibinfo {year}
	{2012}{\natexlab{a}})}\BibitemShut {NoStop}%
\bibitem [{\citenamefont {Neek-Amal}\ and\ \citenamefont
	{Peeters}(2012{\natexlab{b}})}]{Neek-Amal2012b}%
\BibitemOpen
\bibfield  {author} {\bibinfo {author} {\bibfnamefont {M.}~\bibnamefont
		{Neek-Amal}}\ and\ \bibinfo {author} {\bibfnamefont {F.~M.}\ \bibnamefont
		{Peeters}},\ }\href {\doibase 10.1103/PhysRevB.85.195446} {\bibfield
	{journal} {\bibinfo  {journal} {Physical Review B}\ }\textbf {\bibinfo
		{volume} {85}},\ \bibinfo {pages} {195446} (\bibinfo {year}
	{2012}{\natexlab{b}})}\BibitemShut {NoStop}%
\bibitem [{\citenamefont {Schneider}\ \emph {et~al.}(2015)\citenamefont
	{Schneider}, \citenamefont {Faria}, \citenamefont {Viola~Kusminskiy},\ and\
	\citenamefont {Sandler}}]{Schneider2015}%
\BibitemOpen
\bibfield  {author} {\bibinfo {author} {\bibfnamefont {M.}~\bibnamefont
		{Schneider}}, \bibinfo {author} {\bibfnamefont {D.}~\bibnamefont {Faria}},
	\bibinfo {author} {\bibfnamefont {S.}~\bibnamefont {Viola~Kusminskiy}}, \
	and\ \bibinfo {author} {\bibfnamefont {N.}~\bibnamefont {Sandler}},\ }\href
{\doibase 10.1103/PhysRevB.91.161407} {\bibfield  {journal} {\bibinfo
		{journal} {Phys. Rev. B}\ }\textbf {\bibinfo {volume} {91}},\ \bibinfo
	{pages} {161407} (\bibinfo {year} {2015})}\BibitemShut {NoStop}%
\bibitem [{\citenamefont {Carrillo-Bastos}\ \emph {et~al.}(2014)\citenamefont
	{Carrillo-Bastos}, \citenamefont {Faria}, \citenamefont {Latg\'{e}},
	\citenamefont {Mireles},\ and\ \citenamefont
	{Sandler}}]{Carrillo-Bastos2014}%
\BibitemOpen
\bibfield  {author} {\bibinfo {author} {\bibfnamefont {R.}~\bibnamefont
		{Carrillo-Bastos}}, \bibinfo {author} {\bibfnamefont {D.}~\bibnamefont
		{Faria}}, \bibinfo {author} {\bibfnamefont {A.}~\bibnamefont {Latg\'{e}}},
	\bibinfo {author} {\bibfnamefont {F.}~\bibnamefont {Mireles}}, \ and\
	\bibinfo {author} {\bibfnamefont {N.}~\bibnamefont {Sandler}},\ }\href
{\doibase 10.1103/PhysRevB.90.041411} {\bibfield  {journal} {\bibinfo
		{journal} {Physical Review B}\ }\textbf {\bibinfo {volume} {90}},\ \bibinfo
	{pages} {041411} (\bibinfo {year} {2014})}\BibitemShut {NoStop}%
\bibitem [{\citenamefont {Moldovan}\ \emph {et~al.}(2013)\citenamefont
	{Moldovan}, \citenamefont {{Ramezani Masir}},\ and\ \citenamefont
	{Peeters}}]{Moldovan2013}%
\BibitemOpen
\bibfield  {author} {\bibinfo {author} {\bibfnamefont {D.}~\bibnamefont
		{Moldovan}}, \bibinfo {author} {\bibfnamefont {M.}~\bibnamefont {{Ramezani
				Masir}}}, \ and\ \bibinfo {author} {\bibfnamefont {F.~M.}\ \bibnamefont
		{Peeters}},\ }\href {\doibase 10.1103/PhysRevB.88.035446} {\bibfield
	{journal} {\bibinfo  {journal} {Physical Review B}\ }\textbf {\bibinfo
		{volume} {88}},\ \bibinfo {pages} {035446} (\bibinfo {year}
	{2013})}\BibitemShut {NoStop}%
\bibitem [{\citenamefont {Poli}\ \emph {et~al.}(2014)\citenamefont {Poli},
	\citenamefont {Arkinstall},\ and\ \citenamefont {Schomerus}}]{Poli2014}%
\BibitemOpen
\bibfield  {author} {\bibinfo {author} {\bibfnamefont {C.}~\bibnamefont
		{Poli}}, \bibinfo {author} {\bibfnamefont {J.}~\bibnamefont {Arkinstall}}, \
	and\ \bibinfo {author} {\bibfnamefont {H.}~\bibnamefont {Schomerus}},\ }\href
{\doibase 10.1103/PhysRevB.90.155418} {\bibfield  {journal} {\bibinfo
		{journal} {Phys. Rev. B}\ }\textbf {\bibinfo {volume} {90}},\ \bibinfo
	{pages} {155418} (\bibinfo {year} {2014})}\BibitemShut {NoStop}%
\bibitem [{\citenamefont {Landau}\ and\ \citenamefont
	{Lifshitz}(1986)}]{LandauBook}%
\BibitemOpen
\bibfield  {author} {\bibinfo {author} {\bibfnamefont {L.~D.}\ \bibnamefont
		{Landau}}\ and\ \bibinfo {author} {\bibfnamefont {E.~M.}\ \bibnamefont
		{Lifshitz}},\ }\href@noop {} {\emph {\bibinfo {title} {Theory of elasticity,
			third edition: volune 7 (course of theoretical physics)}}}\ (\bibinfo
{publisher} {Springer},\ \bibinfo {year} {1986})\BibitemShut {NoStop}%
\bibitem [{\citenamefont {Goerbig}(2011)}]{Goerbig2011}%
\BibitemOpen
\bibfield  {author} {\bibinfo {author} {\bibfnamefont {M.~O.}\ \bibnamefont
		{Goerbig}},\ }\href {\doibase 10.1103/RevModPhys.83.1193} {\bibfield
	{journal} {\bibinfo  {journal} {Rev. Mod. Phys.}\ }\textbf {\bibinfo {volume}
		{83}},\ \bibinfo {pages} {1193} (\bibinfo {year} {2011})}\BibitemShut
{NoStop}%
\bibitem [{\citenamefont {Suzuura}\ and\ \citenamefont
	{Ando}(2002)}]{Suzuura2002}%
\BibitemOpen
\bibfield  {author} {\bibinfo {author} {\bibfnamefont {H.}~\bibnamefont
		{Suzuura}}\ and\ \bibinfo {author} {\bibfnamefont {T.}~\bibnamefont {Ando}},\
}\href {\doibase 10.1103/PhysRevB.65.235412} {\bibfield  {journal} {\bibinfo
	{journal} {Phys. Rev. B}\ }\textbf {\bibinfo {volume} {65}},\ \bibinfo
{pages} {235412} (\bibinfo {year} {2002})}\BibitemShut {NoStop}%
\bibitem [{\citenamefont {Faria}\ \emph {et~al.}(2013)\citenamefont {Faria},
	\citenamefont {Latg\'{e}}, \citenamefont {Ulloa},\ and\ \citenamefont
	{Sandler}}]{Faria2013}%
\BibitemOpen
\bibfield  {author} {\bibinfo {author} {\bibfnamefont {D.}~\bibnamefont
		{Faria}}, \bibinfo {author} {\bibfnamefont {A.}~\bibnamefont {Latg\'{e}}},
	\bibinfo {author} {\bibfnamefont {S.~E.}\ \bibnamefont {Ulloa}}, \ and\
	\bibinfo {author} {\bibfnamefont {N.}~\bibnamefont {Sandler}},\ }\href
{\doibase 10.1103/PhysRevB.87.241403} {\bibfield  {journal} {\bibinfo
		{journal} {Physical Review B}\ }\textbf {\bibinfo {volume} {87}},\ \bibinfo
	{pages} {241403} (\bibinfo {year} {2013})}\BibitemShut {NoStop}%
\bibitem [{\citenamefont {Pereira}\ \emph {et~al.}(2010)\citenamefont
	{Pereira}, \citenamefont {Ribeiro}, \citenamefont {Peres},\ and\
	\citenamefont {{Castro Neto}}}]{Pereira2010}%
\BibitemOpen
\bibfield  {author} {\bibinfo {author} {\bibfnamefont {V.~M.}\ \bibnamefont
		{Pereira}}, \bibinfo {author} {\bibfnamefont {R.~M.}\ \bibnamefont
		{Ribeiro}}, \bibinfo {author} {\bibfnamefont {N.~M.~R.}\ \bibnamefont
		{Peres}}, \ and\ \bibinfo {author} {\bibfnamefont {A.~H.}\ \bibnamefont
		{{Castro Neto}}},\ }\href {http://stacks.iop.org/0295-5075/92/i=6/a=67001}
{\bibfield  {journal} {\bibinfo  {journal} {EPL (Europhysics Letters)}\
	}\textbf {\bibinfo {volume} {92}},\ \bibinfo {pages} {67001} (\bibinfo {year}
	{2010})}\BibitemShut {NoStop}%
\bibitem [{\citenamefont {Pellegrino}\ \emph {et~al.}(2010)\citenamefont
	{Pellegrino}, \citenamefont {Angilella},\ and\ \citenamefont
	{Pucci}}]{Pellegrino2010}%
\BibitemOpen
\bibfield  {author} {\bibinfo {author} {\bibfnamefont {F.~M.~D.}\
		\bibnamefont {Pellegrino}}, \bibinfo {author} {\bibfnamefont {G.~G.~N.}\
		\bibnamefont {Angilella}}, \ and\ \bibinfo {author} {\bibfnamefont
		{R.}~\bibnamefont {Pucci}},\ }\href {\doibase 10.1103/PhysRevB.81.035411}
{\bibfield  {journal} {\bibinfo  {journal} {Physical Review B}\ }\textbf
	{\bibinfo {volume} {81}},\ \bibinfo {pages} {035411} (\bibinfo {year}
	{2010})}\BibitemShut {NoStop}%
\bibitem [{\citenamefont {Ma\~nes}\ \emph {et~al.}(2013)\citenamefont
	{Ma\~nes}, \citenamefont {de~Juan}, \citenamefont {Sturla},\ and\
	\citenamefont {Vozmediano}}]{Manes2013}%
\BibitemOpen
\bibfield  {author} {\bibinfo {author} {\bibfnamefont {J.~L.}\ \bibnamefont
		{Ma\~nes}}, \bibinfo {author} {\bibfnamefont {F.}~\bibnamefont {de~Juan}},
	\bibinfo {author} {\bibfnamefont {M.}~\bibnamefont {Sturla}}, \ and\ \bibinfo
	{author} {\bibfnamefont {M.~A.~H.}\ \bibnamefont {Vozmediano}},\ }\href
{\doibase 10.1103/PhysRevB.88.155405} {\bibfield  {journal} {\bibinfo
		{journal} {Phys. Rev. B}\ }\textbf {\bibinfo {volume} {88}},\ \bibinfo
	{pages} {155405} (\bibinfo {year} {2013})}\BibitemShut {NoStop}%
\bibitem [{\citenamefont {Masir}\ \emph {et~al.}(2013)\citenamefont {Masir},
	\citenamefont {Moldovan},\ and\ \citenamefont {Peeters}}]{Ramezani2013}%
\BibitemOpen
\bibfield  {author} {\bibinfo {author} {\bibfnamefont {M.~R.}\ \bibnamefont
		{Masir}}, \bibinfo {author} {\bibfnamefont {D.}~\bibnamefont {Moldovan}}, \
	and\ \bibinfo {author} {\bibfnamefont {F.}~\bibnamefont {Peeters}},\ }\href
{\doibase http://dx.doi.org/10.1016/j.ssc.2013.04.001} {\bibfield  {journal}
	{\bibinfo  {journal} {Solid State Communications}\ }\textbf {\bibinfo
		{volume} {175–176}},\ \bibinfo {pages} {76 } (\bibinfo {year}
	{2013})}\BibitemShut {NoStop}%
\bibitem [{\citenamefont {de~Juan}\ \emph {et~al.}(2013)\citenamefont
	{de~Juan}, \citenamefont {Ma\~nes},\ and\ \citenamefont
	{Vozmediano}}]{deJuan2013}%
\BibitemOpen
\bibfield  {author} {\bibinfo {author} {\bibfnamefont {F.}~\bibnamefont
		{de~Juan}}, \bibinfo {author} {\bibfnamefont {J.~L.}\ \bibnamefont
		{Ma\~nes}}, \ and\ \bibinfo {author} {\bibfnamefont {M.~A.~H.}\ \bibnamefont
		{Vozmediano}},\ }\href {\doibase 10.1103/PhysRevB.87.165131} {\bibfield
	{journal} {\bibinfo  {journal} {Phys. Rev. B}\ }\textbf {\bibinfo {volume}
		{87}},\ \bibinfo {pages} {165131} (\bibinfo {year} {2013})}\BibitemShut
{NoStop}%
\bibitem [{\citenamefont {Jang}\ \emph {et~al.}(2014)\citenamefont {Jang},
	\citenamefont {Kim}, \citenamefont {Shin}, \citenamefont {Wang},
	\citenamefont {Jang}, \citenamefont {Kim}, \citenamefont {Lee}, \citenamefont
	{Kim}, \citenamefont {Song},\ and\ \citenamefont {Kahng}}]{Jang2014}%
\BibitemOpen
\bibfield  {author} {\bibinfo {author} {\bibfnamefont {W.-J.}\ \bibnamefont
		{Jang}}, \bibinfo {author} {\bibfnamefont {H.}~\bibnamefont {Kim}}, \bibinfo
	{author} {\bibfnamefont {Y.-R.}\ \bibnamefont {Shin}}, \bibinfo {author}
	{\bibfnamefont {M.}~\bibnamefont {Wang}}, \bibinfo {author} {\bibfnamefont
		{S.~K.}\ \bibnamefont {Jang}}, \bibinfo {author} {\bibfnamefont
		{M.}~\bibnamefont {Kim}}, \bibinfo {author} {\bibfnamefont {S.}~\bibnamefont
		{Lee}}, \bibinfo {author} {\bibfnamefont {S.-W.}\ \bibnamefont {Kim}},
	\bibinfo {author} {\bibfnamefont {Y.~J.}\ \bibnamefont {Song}}, \ and\
	\bibinfo {author} {\bibfnamefont {S.-J.}\ \bibnamefont {Kahng}},\ }\href
{\doibase http://dx.doi.org/10.1016/j.carbon.2014.03.015} {\bibfield
	{journal} {\bibinfo  {journal} {Carbon}\ }\textbf {\bibinfo {volume} {74}},\
	\bibinfo {pages} {139 } (\bibinfo {year} {2014})}\BibitemShut {NoStop}%
\bibitem [{\citenamefont {Li}\ \emph {et~al.}(2015)\citenamefont {Li},
	\citenamefont {Bai}, \citenamefont {Yin}, \citenamefont {Qiao}, \citenamefont
	{Wang},\ and\ \citenamefont {He}}]{Li2015}%
\BibitemOpen
\bibfield  {author} {\bibinfo {author} {\bibfnamefont {S.-Y.}\ \bibnamefont
		{Li}}, \bibinfo {author} {\bibfnamefont {K.-K.}\ \bibnamefont {Bai}},
	\bibinfo {author} {\bibfnamefont {L.-J.}\ \bibnamefont {Yin}}, \bibinfo
	{author} {\bibfnamefont {J.-B.}\ \bibnamefont {Qiao}}, \bibinfo {author}
	{\bibfnamefont {W.-X.}\ \bibnamefont {Wang}}, \ and\ \bibinfo {author}
	{\bibfnamefont {L.}~\bibnamefont {He}},\ }\href@noop {} {\  (\bibinfo {year}
	{2015})},\ \Eprint {http://arxiv.org/abs/11506.07965} {arXiv:11506.07965}
\BibitemShut {NoStop}%
\bibitem [{\citenamefont {Beenakker}(2008)}]{Beenakker2008}%
\BibitemOpen
\bibfield  {author} {\bibinfo {author} {\bibfnamefont {C.~W.~J.}\
		\bibnamefont {Beenakker}},\ }\href {\doibase 10.1103/RevModPhys.80.1337}
{\bibfield  {journal} {\bibinfo  {journal} {Rev. Mod. Phys.}\ }\textbf
	{\bibinfo {volume} {80}},\ \bibinfo {pages} {1337} (\bibinfo {year}
	{2008})}\BibitemShut {NoStop}%
\bibitem [{Note1()}]{Note1}%
\BibitemOpen
\bibinfo {note} {We note that the $c_1$ and $c_2$ coefficients contain sign
	factors depending on both valley and band index \cite {Goerbig2011}. In the
	discussions of Section~\ref {sec:analyticA}, however, we omit these factors
	for simplicity since we focus on features like the LDOS which only depend on
	the magnitude of the wavefunction.}\BibitemShut {Stop}%
\bibitem [{\citenamefont {Juan}\ \emph {et~al.}(2011)\citenamefont {Juan},
	\citenamefont {Cortijo}, \citenamefont {Vozmediano},\ and\ \citenamefont
	{Cano}}]{Juan2011}%
\BibitemOpen
\bibfield  {author} {\bibinfo {author} {\bibfnamefont {F.~D.}\ \bibnamefont
		{Juan}}, \bibinfo {author} {\bibfnamefont {A.}~\bibnamefont {Cortijo}},
	\bibinfo {author} {\bibfnamefont {M.~A.~H.}\ \bibnamefont {Vozmediano}}, \
	and\ \bibinfo {author} {\bibfnamefont {A.}~\bibnamefont {Cano}},\ }\href
{\doibase 10.1038/nphys2034} {\bibfield  {journal} {\bibinfo  {journal}
		{Nature Physics}\ }\textbf {\bibinfo {volume} {7}},\ \bibinfo {pages} {810}
	(\bibinfo {year} {2011})}\BibitemShut {NoStop}%
\bibitem [{\citenamefont {Zabet-Khosousi}\ \emph {et~al.}(2014)\citenamefont
	{Zabet-Khosousi}, \citenamefont {Zhao}, \citenamefont {Pálová},
	\citenamefont {Hybertsen}, \citenamefont {Reichman}, \citenamefont
	{Pasupathy},\ and\ \citenamefont {Flynn}}]{Zabet2014}%
\BibitemOpen
\bibfield  {author} {\bibinfo {author} {\bibfnamefont {A.}~\bibnamefont
		{Zabet-Khosousi}}, \bibinfo {author} {\bibfnamefont {L.}~\bibnamefont
		{Zhao}}, \bibinfo {author} {\bibfnamefont {L.}~\bibnamefont {Pálová}},
	\bibinfo {author} {\bibfnamefont {M.~S.}\ \bibnamefont {Hybertsen}}, \bibinfo
	{author} {\bibfnamefont {D.~R.}\ \bibnamefont {Reichman}}, \bibinfo {author}
	{\bibfnamefont {A.~N.}\ \bibnamefont {Pasupathy}}, \ and\ \bibinfo {author}
	{\bibfnamefont {G.~W.}\ \bibnamefont {Flynn}},\ }\href {\doibase
	10.1021/ja408463g} {\bibfield  {journal} {\bibinfo  {journal} {Journal of the
			American Chemical Society}\ }\textbf {\bibinfo {volume} {136}},\ \bibinfo
	{pages} {1391} (\bibinfo {year} {2014})}\BibitemShut {NoStop}%
\bibitem [{\citenamefont {Power}\ and\ \citenamefont
	{Ferreira}(2011)}]{Power2011}%
\BibitemOpen
\bibfield  {author} {\bibinfo {author} {\bibfnamefont {S.~R.}\ \bibnamefont
		{Power}}\ and\ \bibinfo {author} {\bibfnamefont {M.~S.}\ \bibnamefont
		{Ferreira}},\ }\href {\doibase 10.1103/PhysRevB.83.155432} {\bibfield
	{journal} {\bibinfo  {journal} {Physical Review B}\ }\textbf {\bibinfo
		{volume} {83}},\ \bibinfo {pages} {155432} (\bibinfo {year}
	{2011})}\BibitemShut {NoStop}%
\bibitem [{\citenamefont {Verbiest}\ \emph {et~al.}(2015)\citenamefont
	{Verbiest}, \citenamefont {Brinker},\ and\ \citenamefont
	{Stampfer}}]{Verbiest2015}%
\BibitemOpen
\bibfield  {author} {\bibinfo {author} {\bibfnamefont {G.~J.}\ \bibnamefont
		{Verbiest}}, \bibinfo {author} {\bibfnamefont {S.}~\bibnamefont {Brinker}}, \
	and\ \bibinfo {author} {\bibfnamefont {C.}~\bibnamefont {Stampfer}},\ }\href
{\doibase 10.1103/PhysRevB.92.075417} {\bibfield  {journal} {\bibinfo
		{journal} {Phys. Rev. B}\ }\textbf {\bibinfo {volume} {92}},\ \bibinfo
	{pages} {075417} (\bibinfo {year} {2015})}\BibitemShut {NoStop}%
\bibitem [{\citenamefont {Settnes}\ \emph
	{et~al.}(2015{\natexlab{b}})\citenamefont {Settnes}, \citenamefont {Power},
	\citenamefont {Lin}, \citenamefont {Petersen},\ and\ \citenamefont
	{Jauho}}]{Settnes2015b}%
\BibitemOpen
\bibfield  {author} {\bibinfo {author} {\bibfnamefont {M.}~\bibnamefont
		{Settnes}}, \bibinfo {author} {\bibfnamefont {S.~R.}\ \bibnamefont {Power}},
	\bibinfo {author} {\bibfnamefont {J.}~\bibnamefont {Lin}}, \bibinfo {author}
	{\bibfnamefont {D.~H.}\ \bibnamefont {Petersen}}, \ and\ \bibinfo {author}
	{\bibfnamefont {A.-P.}\ \bibnamefont {Jauho}},\ }\href
{http://arxiv.org/abs/1506.06509} {\  (\bibinfo {year}
	{2015}{\natexlab{b}})},\ \Eprint {http://arxiv.org/abs/1506.06509}
{arXiv:1506.06509} \BibitemShut {NoStop}%
\bibitem [{\citenamefont {{Ramezani Masir}}\ \emph {et~al.}(2009)\citenamefont
	{{Ramezani Masir}}, \citenamefont {Matulis},\ and\ \citenamefont
	{Peeters}}]{RamezaniMasir2009}%
\BibitemOpen
\bibfield  {author} {\bibinfo {author} {\bibfnamefont {M.}~\bibnamefont
		{{Ramezani Masir}}}, \bibinfo {author} {\bibfnamefont {A.}~\bibnamefont
		{Matulis}}, \ and\ \bibinfo {author} {\bibfnamefont {F.~M.}\ \bibnamefont
		{Peeters}},\ }\href {\doibase 10.1103/PhysRevB.79.155451} {\bibfield
	{journal} {\bibinfo  {journal} {Physical Review B}\ }\textbf {\bibinfo
		{volume} {79}},\ \bibinfo {pages} {155451} (\bibinfo {year}
	{2009})}\BibitemShut {NoStop}%
\bibitem [{\citenamefont {{Ramezani Masir}}\ \emph {et~al.}(2011)\citenamefont
	{{Ramezani Masir}}, \citenamefont {Vasilopoulos},\ and\ \citenamefont
	{Peeters}}]{RamezaniMasir2011}%
\BibitemOpen
\bibfield  {author} {\bibinfo {author} {\bibfnamefont {M.}~\bibnamefont
		{{Ramezani Masir}}}, \bibinfo {author} {\bibfnamefont {P.}~\bibnamefont
		{Vasilopoulos}}, \ and\ \bibinfo {author} {\bibfnamefont {F.~M.}\
		\bibnamefont {Peeters}},\ }\href {\doibase 10.1088/0953-8984/23/31/315301}
{\bibfield  {journal} {\bibinfo  {journal} {Journal of physics. Condensed
			matter : an Institute of Physics journal}\ }\textbf {\bibinfo {volume}
		{23}},\ \bibinfo {pages} {315301} (\bibinfo {year} {2011})}\BibitemShut
{NoStop}%
\end{thebibliography}

%

\end{document}